\documentclass[10pt]{article}

\usepackage{epsf}
\usepackage{epsfig}
\usepackage{amsmath}
\usepackage{amsthm}
\usepackage{amssymb}
\usepackage{amscd}
\usepackage{rotating}
\usepackage{wrapfig}
\usepackage{psfrag}

\def\1I{\relax{\rm 1\kern-.25em \rm l}} % Operator-Eins

\newcommand{\unity}{\1I}

\newcommand{\ROrder}{\raisebox{-1.75ex}{$\rightarrow$}\hspace{-2ex}}
\newcommand{\LOrder}{\raisebox{-1.75ex}{$\leftarrow$}\hspace{-2ex}}
\newcommand{\MyIm}{\Im\mathfrak{m}}
\newcommand{\MyRe}{\Re\mathfrak{e}}

\theoremstyle{plain}

\newtheorem{defn}{Definition}

\newtheorem{lem}{Lemma}
\newtheorem{prop}{Proposition}

\newtheorem{ex}{Example}

\theoremstyle{remark}
\newtheorem{rem}{Remark}
\font\figurefont=cmbx10 scaled 1100

\def\href#1#2{#2}

\begin{document}

\thispagestyle{empty}
\vspace{2truecm}
\centerline{\bf \Large Review of SU(2)-Calibrations}
\vspace{.5truecm}

\newcounter{Institut}
\vspace{1.5truecm}
\centerline{\bf Andr\'e Miemiec}

\vskip1cm
\parbox{.95\textwidth}{
\centerline{Institut f\"ur Physik~~}
\centerline{Humboldt Universit\"at}
\centerline{D-12489 Berlin, Germany}
\centerline{Newtonstr. 15}
\centerline{miemiec@physik.hu-berlin.de}       
}

\vspace{.4truecm}

\vspace{1.0truecm}
%%%%%%%%%%%%%%%%%%%%%%%%%%%%%%%%%%%%%%%%%%%%%%%%%%%%%%%%%%%%%
%      ABSTRACT
%%%%%%%%%%%%%%%%%%%%%%%%%%%%%%%%%%%%%%%%%%%%%%%%%%%%%%%%%%%%%
%\vspace{.4truecm}
\begin{abstract}
\noindent
The purpose of this article is to provide a review of 
SU(2)-calibrations. The focus is on developing all 
techniques in full detail by studying selected examples.  
The supergravity point of view and  the string theoretic 
one are explained. 
\end{abstract}
%%%%%%%%%%%%%%%%%%%%%%%%%%%%%%%%%%%%%%%%%%%%%%%%%%%%%%%%%%%%%
\bigskip \bigskip
\newpage

\tableofcontents

\newpage

%%%%%%%%%%%%%%%%%%%%%%%%%%%%%%%%%%%%%%%%%%%%%%%%%%%%%%%%%%%%%%%%%
% Section: Introduction
%%%%%%%%%%%%%%%%%%%%%%%%%%%%%%%%%%%%%%%%%%%%%%%%%%%%%%%%%%%%%%%%%

\setcounter{page}{1}
\section{Introduction}

In this notes we would like to give a self contained introduction into 
the subject of calibrated geometries  \cite{HL} while setting the focus on
the important example of $SU(2)$-calibrated geometries. Calibrated  
geometries arise in supergravity and string theory quite 
naturally in the investigation of 
classical solutions of the equations of motion (eom), which preserve 
some amount of supersymmetry\footnote
%%%%%%%%%%%%%%%%%%%%%%%%%%%%%%%%%%%%%%%%%%%%%%%%%%%%%%%%%%%%%%%%
% FOOTNOTE
%%%%%%%%%%%%%%%%%%%%%%%%%%%%%%%%%%%%%%%%%%%%%%%%%%%%%%%%%%%%%%%%
{
  For the case of 11d supergravity 
  some important examples of brane solutions are discussed
  in \cite{Miemiec:2005ry}.
}
%%%%%%%%%%%%%%%%%%%%%%%%%%%%%%%%%%%%%%%%%%%%%%%%%%%%%%%%%%%%%%%%
% FOOTNOTE
%%%%%%%%%%%%%%%%%%%%%%%%%%%%%%%%%%%%%%%%%%%%%%%%%%%%%%%%%%%%%%%%
(Killing spinors) \cite{Becker:1995kb,Witten:1997sc,Smith}. 
Typical examples are provided by the widely used 
brane solutions \cite{Gauntlett:1998vk,Gauntlett:1998wb,Barwald:1999ux,Lust:1999pq}. 
Nowadays the application of calibrated geometries 
is one of the main computational tools in supergravity and string
theory and the objective of current research. Therefore an
understanding of the basic concepts is very useful \cite{Acharya:1998yv,Acharya:1998en,Gibbons:1998hm}. In recent
years the attempts to provide a full classification of supersymmetric 
solutions of  all supergravities of physical interest have benefited a
lot from the progress in understanding (generalised) 
calibrations \cite{Gutowski:1999iu,Gutowski:1999tu,Gauntlett:2002fz,Figueroa-O'Farrill:2002ft,Gauntlett:2003fk}. Further work in this direction
may be performed and belongs certainly to the most inspiring developments 
in supergravity and string theory, nowadays.
To set the scene we want to motivate the origin of calibrated
geometries by giving a simple example. 
For this purpose we consider branes, which 
prove to be generic supersymmetric solutions of supergravities and are
most important in applications. 
One can also look at them from a different point of view. Rather
then considering them as solutions of a given supergravity one might
regard them as dynamical objects moving in a fixed supergravity 
background, too. The action from which the
dynamics of the brane in the background follows 
is called p-brane action \cite{Fradkin:1985qd,Leigh:1989jq,Witten:1996im}.

\vspace{-1ex}
\begin{ex}\label{BSP1}
{\rm
In a flat background and in the absence of other gauge fields 
the p-brane action  governing the dynamics is simply the volume 
functional of the brane embedded into  flat D-dimensional spacetime. 
The solutions of the corresponding equations of motion are minimal 
surfaces. If $f:M^{1,p}\rightarrow {\mathbb{R}}^{1,D-1}$ is the embedding map
and for simplicity we assume it  describes effectively the embedding 
of a 2-manifold ${\mathcal{C}}$ into ${\mathbb{R}}^4$, i.e.\\ 
\begin{eqnarray*}
    {\mathbb{R}}^{1,p-2}\times \underline{\,{\mathcal{C}}}
    &\longrightarrow& {\mathbb{R}}^{1,p-2}\times
    \underline{\,{\mathbb{R}}^4}\times{\mathbb{R}}^{D-p-3}~,\\
    (x^0,x^1,\ldots, x^{p-2},\underline{\xi^1,\xi^2})&\mapsto&(x^0,x^1,\ldots,
    x^{p-2},\underline{\xi^1,\xi^2,X^1,X^2},\ldots,x^{D-1})
\end{eqnarray*}\\
with $X^1=X^1(\xi^1,\xi^2)$ and $X^2=X^2(\xi^1,\xi^2)$. We can
consider ${{\mathbb{R}}^4}$ as the complex space ${{\mathbb{C}}^2}$
by introducing the following complex coordinates\\ 
\parbox{\textwidth}
{\hspace{0.5cm}
 \parbox{0.45\textwidth}
 {
      %%%%%%%%%%%%%%%%%%%%%%%%%%%%%%%%%%%%%%%%%%%%%%%
      %    The three independent hypersurfaces
      %%%%%%%%%%%%%%%%%%%%%%%%%%%%%%%%%%%%%%%%%%%%%%%
      \begin{center}
      \psfrag{r1}{$\vec{\mathfrak{r}}_1$}
      \psfrag{r2}{$\vec{\mathfrak{r}}_2$}
      \psfrag{C}{$\mathcal{C}$}
         \mbox{ \begin{turn}{0}%
           \epsfig{file=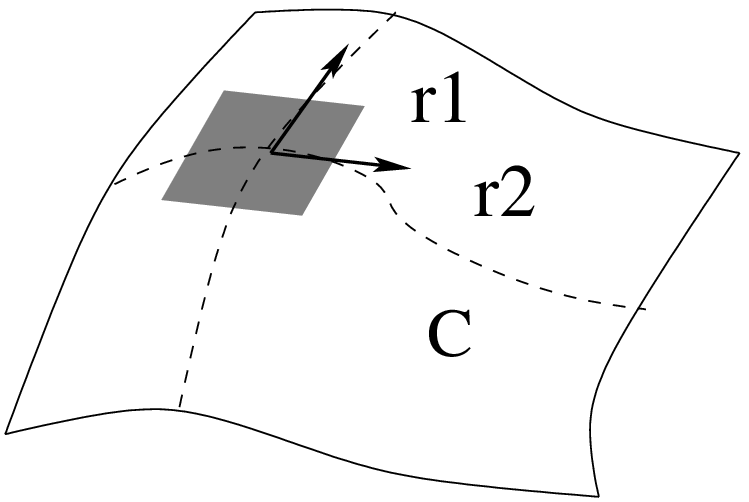,width=6cm}
                \end{turn}
         }
      \end{center}
      %%%%%%%%%%%%%%%%%%%%%%%%%%%%%%%%%%%%%%%%%%%%%%%
 }\hfill
 \parbox{0.45\textwidth}
 {
    \begin{eqnarray*}
              {z^1} &=& {\xi^1}\,+\,i\,{X^1}\\[1ex]
              {z^2} &=& {\xi^2}\,+\,i\,{X^2}~.
      \end{eqnarray*}
      ${\mathbb{C}}^2$ is a K\"ahler manifold with K\"ahler form 
      \begin{equation}\label{Kaehler}
        {\omega} ~=~ \frac{i}{2}\left(\,
                                       dz^1\wedge d\bar{z}^1 ~+~
                                       dz^2\wedge d\bar{z}^2
                              \right)~.
      \end{equation}
      In addition one can define a second closed form via  
      \begin{equation}\label{ComplVolume}
        {\Omega} ~=~ dz^1\wedge dz^2~.  
      \end{equation}
 }\hspace{0.5cm}
}
The induced metric on ${\mathcal{C}}$ is computed in the standard way
by computing all possible scalar products of the tangent vectors
$\vec{\mathfrak{r}}_1$ and $\vec{\mathfrak{r}}_2$ at a 
point $(\xi^1,\xi^2)$ and we obtain:\\ 
\begin{eqnarray}\label{Eq_induced_metric}
      (f^{\ast}g)_{ij} &=& \left(
                       \begin{array}{cc}
                          1+\left(\partial_1 X_1\right)^2
                           +\left(\partial_1 X_2\right)^2&
                           \partial_1 X_1\partial_2 X_1
                          +\partial_1 X_2\partial_2 X_2\\[0.5ex]
                           \partial_1 X_1\partial_2 X_1
                          +\partial_1 X_2\partial_2 X_2 &
                          1+\left(\partial_2 X_1\right)^2
                           +\left(\partial_2 X_2\right)^2
                       \end{array}
                 \right)~.           
\end{eqnarray}\\
The volume functional expressed in terms of the determinant of the  
induced metric $(f^\ast g)_{ij}$ on ${\mathcal{C}}$ reads 
(see appendix \ref{Sec_Det2dMetrik})\\   
\begin{eqnarray*}
     S ~=~ \int\limits_{{\mathbb{R}}^{1,p-2}\times {\mathcal{C}}^2} 
           d^{p+1}x\;\sqrt{|g|}
       &=& {\rm Vol}({\mathbb{R}}^{1,p-2})\cdot
           \int\limits_{{\mathcal{C}}^2}d^2x\;\sqrt{|f^\ast g|}\nonumber\\
       &=& {\rm Vol}({\mathbb{R}}^{1,p-2})\cdot
           \int\limits_{{\mathcal{C}}^2}d^2x\; 
                                         \sqrt{\,\vbox{\vspace{2.5ex}}
                                               [f^\ast\omega]^2 \,+\,
                                               [f^\ast\MyRe\,\Omega]^2 \,+\,
                                               [f^\ast\MyIm\,\Omega]^2\,
                                              }\nonumber\\
       &\geq& {\rm Vol}({\mathbb{R}}^{1,p-2})\cdot
           \int\limits_{{\mathcal{C}}^2}  f^\ast\omega~.
\end{eqnarray*}\\
The volume is bounded from below. 
If the bound is saturated, i.e. the equality sign holds, one obtains 
a peculiar geometry given by the equation}\\ 
\begin{eqnarray}
    f^\ast\omega &=& \sqrt{|g|}\;d\xi^1\wedge d\xi^2\,,
\end{eqnarray}\\  
{\rm i.e. the pullback of the K\"ahler form $\omega$ is the volume form on
${\mathcal{C}}$. This statement is a local one. 
According to Wirtingers theorem this is a 
property characteristic of complex submanifolds 
of ${\mathbb{C}}^2$ \cite{PrOfAlGe}. So the surfaces calibrated with
respect to the K\"ahler form $\omega$ are just the complex
submanifolds of ${\mathbb{C}}^2$. Similar bounds but formally 
different geometries can be established 
by utilising one of the other three squares in $|f^\ast g|$, too}\footnote 
%%%%%%%%%%%%%%%%%%%%%%%%%%%%%%%%%%%%%%%%%%%%%%%%%%%%%%%%%%%%%%%
%  FOOTNOTE
%%%%%%%%%%%%%%%%%%%%%%%%%%%%%%%%%%%%%%%%%%%%%%%%%%%%%%%%%%%%%%%
{
 This seems to be ambiguous. A complete
 resolution of this ambiguity will be given in 
 this review.
}.\\[-5ex]
%%%%%%%%%%%%%%%%%%%%%%%%%%%%%%%%%%%%%%%%%%%%%%%%%%%%%%%%%%%%%%%
%  FOOTNOTE
%%%%%%%%%%%%%%%%%%%%%%%%%%%%%%%%%%%%%%%%%%%%%%%%%%%%%%%%%%%%%%%
\flushright{$\Box$}
\end{ex}
\begin{defn}{\bf (Preliminary) :} A calibration consists of a closed 
differential form $\omega$ , which provides a lower bound for 
the volume of a brane solution such that for all submanifolds $\mathcal{C}$ 
\begin{eqnarray}\label{calform}
    {\rm d}vol_{\mathcal{C}} \ge f^\ast \omega\left.\vbox{\vspace{2.5ex}}\right|_{T\mathcal{C}}\ ,
\end{eqnarray}
when evaluated on any tangent plane. If a submanifold (brane) 
saturates the bound we call it calibrated. The saturation of 
the bound can be translated into constraints on the geometry of 
the submanifold (brane).
\end{defn}

\noindent
In the next sections we want to develop an understanding 
for the interplay of  closed differential forms defining 
a calibration and geometrical constraints.
In order to be as explicit as possible we limit ourself to a detailed  
study of the simplest but nevertheless nontrivial example of 
$SU(2)$-calibrated geometry. The reason for calling it a $SU(2)$-
calibrated geometry will become clear later. 
The advantage of  $SU(2)$-calibrations is, that everything can be 
computed easily and so most insight into the subject is possible.
It can be seen as a role model for more sophisticated 
constructions of calibrated geometries \cite{HL}. 
There exists other reviews, which might be
consulted for further reading \cite{Figueroa-O'Farrill:1998su,Townsend:1999hi}.   
Here we choose the following strategy. 
As was pointed out 
in example \ref{BSP1} the study of the local geometry is enough 
in order to derive the constraints (differential equations) 
governing the geometry of the manifold.
Therefore our plan is to start with a 
formal discussion of the local geometry of $SU(2)$-calibrated submanifolds 
by studying the geometry of two planes 
in $\mathbb{R}^4$ in section \ref{Sec_G24}, 
first \cite{Souriau,GeometricAsymptotics,Woodhouse}. 
Having achieved a complete understanding of the geometries 
we want to investigate the whole topic from the perspective 
often taken in string theory in section \ref{Sec_StringPict} 
\cite{Berkooz:1996km,Ohta:1997fr,Townsend:1997mk,Witten:2000mf,Mitra:2000wr,Ohta:2001dh}.
Here we rederive the $SU(2)$-conditions from the string picture 
in the case of an D2-D4 bound state \cite{KLM}.
Finally we generalise the recipe to generate the differential
equations to the case of $SU(d)$-calibrations in section
\ref{Sec_SLAC_D} \cite{Karch:1998sj,PhD}.

\section{Grassmanians of 2-planes}
\label{Sec_G24}

\noindent
\subsection{Grassmanian $G(2,4)$}

\noindent
The Grassmanian of 2-planes in ${\mathbb{R}}^4$ is defined through:\\
\begin{eqnarray*}
         G(2,4)~=~\{\,\xi\wedge\eta\,|\,\xi,\,\eta\in
                     \Omega^1({\mathbb{R}}^4)\,\} 
               ~\subset~ \Lambda^2{\mathbb{R}}^4, 
\end{eqnarray*}\\ 
i.e. as a subset of the six dimensional space $\Lambda^2{\mathbb{R}}^4$.
Its elements are 2-planes generated by two vectors dual to the two 
1-forms $\xi$ and $\eta$.\\

\noindent
We would like to study the structure of the manifold $G(2,4)$. To that 
purpose we consider the action of the Hodge star operator 
on $\Lambda^2{\mathbb{R}}^4$. The action of the Hodge star is simply 
defined by\\
\begin{eqnarray*}
   \ast\,(\,dx^i\wedge dx^j\,) ~=~ \hbox{sgn}(\,i\,j\,k\,l\,)\,dx^k\wedge dx^l.
\end{eqnarray*}\\
Since $\ast^2=\unity$ there is a split of the six dimensional space\\     
\begin{eqnarray*}
   \Lambda^2{\mathbb{R}}^4~=~\Lambda_{+}^{2}{\mathbb{R}}^4\oplus
                             \Lambda_{-}^{2}{\mathbb{R}}^4
\end{eqnarray*}\\
into three dimensional eigenspaces of $\ast$. The split is realized
via the projections 
$P_{\pm} ~=~ \frac{1}{2}\,\left(\,1\,\pm\,\ast\,\right)$.\\
%%%%%%%%%%%%%%%%%%%%%%%%%%%%%%%%%%%%%%%%%%%%%%%%%%%%%%%%%%%%%%%%%%%%%%%%%%%
%                     \Lambda^2{\mathbb{R}}^4
%                       /     /   \       \
%            P_{+}    /     /       \       \    P_{-}
%                    |    / i      j  \      |
%                   \/  /               \    \/
%  \Lambda_{+}^2{\mathbb{R}}^4      \Lambda_{-}^2{\mathbb{R}}^4
%
%%%%%%%%%%%%%%%%%%%%%%%%%%%%%%%%%%%%%%%%%%%%%%%%%%%%%%%%%%%%%%%%%%%%%%%%%%
\begin{displaymath}\begin{CD}
         &             &\Lambda^2{\mathbb{R}}^4&           &\\
   P_{+} &\swarrow     &                       &\searrow   & P_{-}\\
   \Lambda_{+}^2{\mathbb{R}}^4 & & & & \Lambda_{-}^2{\mathbb{R}}^4 \\
\end{CD}\end{displaymath}\\
%%%%%%%%%%%%%%%%%%%%%%%%%%%%%%%%%%%%%%%%%%%%%%%%%%%%%%%%%%%%%%%%%%%%%%%%%% 

\noindent
Now we investigate the decomposition of $G(2,4)$ according to 
this split.
Every element $x$ of $G(2,4)$ is of the form $x=\xi\wedge\eta$, 
where the vectors $\xi$ and $\eta$ can be chosen to be orthonormal.  
The following statement is true:\\
\begin{prop}  
\begin{eqnarray*}
    x \in G(2,4),\hbox{ iff }   x\wedge x ~=~ 0\,.   
\end{eqnarray*}\\[-2ex]
\end{prop} 
\begin{proof}
{\bf Without proof.}
\end{proof}

\begin{prop}  
\begin{eqnarray*}
    G(2,4) &=& S^2_{+}\times S^2_{-} 
\end{eqnarray*}\\[-2ex]
\end{prop} 
\begin{proof}

\noindent
Each $x\in G(2,4)$ can be decomposed according to\\  
\begin{eqnarray*}
           x &=& P_{+}\,x\, +\, P_{-}\,x 
\end{eqnarray*}\\
and we will see that this splitting defines a map of 
$G(2,4)$ into $S_{+}^2\times S_{-}^2$. To that purpose 
we compute the norm of the three vector obtained via 
the projection of $x$ into the ``+''-eigenspace of $\ast$. 
The following equation holds\footnote
%%%%%%%%%%%%%%%%%%%%%%%%%%%%%%%%%%%%%%%%%%%%%%%%%%%%%%%%%%%%%%%%
% FOOTNOTE
%%%%%%%%%%%%%%%%%%%%%%%%%%%%%%%%%%%%%%%%%%%%%%%%%%%%%%%%%%%%%%%%
{
  This is due to the definition of the scalar product of two 
  differential p-forms 
  $\omega\,\wedge\,\ast\,\eta = <\omega,\,\eta>\,{\rm vol}$. Here 
  {\rm vol} denotes the volume form.
}:\\ 
%%%%%%%%%%%%%%%%%%%%%%%%%%%%%%%%%%%%%%%%%%%%%%%%%%%%%%%%%%%%%%%%
% FOOTNOTE
%%%%%%%%%%%%%%%%%%%%%%%%%%%%%%%%%%%%%%%%%%%%%%%%%%%%%%%%%%%%%%%%
\begin{eqnarray*}
   \|P_{+}x\|^2\,{\rm vol}&=&  P_{+}x \wedge \ast P_{+}x 
                          ~=~P_{+}x \wedge P_{+}x\\
                        &=& \frac{1}{2}\,\left(\,x\,+\,\ast\,x\,\right)
                     \wedge \frac{1}{2}\,\left(\,x\,+\,\ast\,x\,\right)\\
                        &=& \frac{1}{2}\,x\,\wedge\,\ast\,x
                        ~=~ \frac{1}{2}\, {\underbrace{\|x\|}_{1}}^2\,
                            {\rm vol}.
\end{eqnarray*}\\
Setting $\|x\|^2=1$ merely reflects the fact that the area
spanned by normed vectors $\xi$ and $\eta$ is equal to one. 
We obtain the result that the three ``vector'' $P_{+}x$ is of norm 1/2.\\

\noindent
The general element of $P_{+} G(2,4)$ can be written as the three ``vector''\\
\begin{eqnarray*}
    P_{+} x &=& a_{12} P_{+}\left(dx^1\wedge dx^2\right)
              + a_{13} P_{+}\left(dx^1\wedge dx^3\right)
              + a_{23} P_{+}\left(dx^2\wedge dx^3\right).
\end{eqnarray*}\\
Using the following  abbreviation
$(a_{12},a_{13},a_{23})\mapsto(a_1,a_2,a_3)=\vec{a}$ the 
following identity holds:\\
\begin{eqnarray*}
   <P_{+}x,\,P_{+}y>\,{\rm vol} &=& \frac{1}{2}\,<\vec{a}\,,\,\vec{b}>\,{\rm vol}.
\end{eqnarray*}\\
Comparing the previous result of the norm of $P_{+}x$ with the latter
identity one obtains\\
\begin{eqnarray*}
    \|P_{+}x\|^2\,{\rm vol} &=& \frac{1}{2}\, \|\,\vec{a}\,\|^{\,2}\,{\rm
                         vol}
                         ~=~  \frac{1}{2}\, {\rm vol}, 
\end{eqnarray*}\\
i.e. the  vector $\vec{a}$ associated with $P_{+}x$ is of unit norm.
A completely analogous result holds for the ``-'' eigenspace:\\
\begin{eqnarray*}
   <P_{-}x,\,P_{-}y>\,{\rm vol} &=& \frac{1}{2}\,<\vec{a}\,,\,\vec{b}>\,{\rm vol}\\
   \|P_{-}x\|^2\,{\rm vol} &=& \frac{1}{2}\, {\underbrace{\|\vec{a}\|}_{1}}^2\, 
                             {\rm vol}
                           ~=~ \frac{1}{2}\, {\underbrace{\|x\|}_{1}}^2\, 
                               {\rm vol}.
\end{eqnarray*}\\
Basically the vector $\vec{a}$ associated with $P_{-}x$ is again of unit norm.
In addition for two 2-forms belonging to the ``+'' and the ``-''
eigenspaces, respectively,  the 
formula\\ 
\begin{eqnarray*}
    \omega_{+} \wedge \omega_{-} &=& 0
\end{eqnarray*}\\
holds, i.e. the decomposition is an orthogonal one.
We obtain the announced map:\\
\begin{eqnarray*}
      \phi: G(2,4) &\longrightarrow& S^2_{+}\times S^2_{-}.\\
              x      &\mapsto&  P_{+}\,x\, +\, P_{-}\,x
\end{eqnarray*}\\

\noindent 
\begin{lem}
The map $\phi: G(2,4)~\longrightarrow~ S^2_{+}\times S^2_{-}$ is bijective.
\end{lem}

\begin{proof}$\left.\right.$\\[0.3cm]
\underline{Surjectivity:}  Choose $(a,b)\in S_{+}^2\times S_{-}^2$
                           and define a section $s$ of $G(2,4)$ by\\
                           \begin{eqnarray*}
                             s:S^2_{+}\times S^2_{-}&\longrightarrow&G(2,4)\\
                                  (a,b)&\mapsto& x~=~\omega_a+\omega_b
                           \end{eqnarray*}\\
                           Then compute\\ 
                           \begin{eqnarray*}
                              x\wedge x&=& \underbrace{
                                              \omega_a\wedge\omega_a
                                           }_{1/2\,{\rm vol}}
                                          +2\underbrace{
                                              \omega_a\wedge\omega_b
                                           }_{0}
                                           +\underbrace{
                                              \omega_b\wedge\omega_b
                                            }_{-1/2\,{\rm vol}} ~=~ 0,
                           \end{eqnarray*}\\
i.e. $x\in G(2,4)$, indeed.\\[2ex]
\underline{Injectivity:} Choose two $x,x'\in G(2,4)$ having the 
                         same images in $S^2_{+}\times S^2_{-}$. 
                         Each can be 
                         written as $x~=~P_{+}x\,+\,P_{-}x$
                         and $x'~=~P_{+}x'\,+\,P_{-}x'$, respectively. 
                         Since the images coincide, i.e.\\ 
                         \begin{eqnarray*}
                            P_{+}(x) &=& P_{+}(x') \\
                            P_{-}(x) &=& P_{-}(x') 
                         \end{eqnarray*}\\  
                         or more detailed\\ 
                         \begin{eqnarray*}
                            x-x' &=& -*(x-x') \\
                            x-x' &=& +*(x-x'), 
                         \end{eqnarray*}\\
                         one concludes $x=x'$ due to the orthogonality of the
                         split, i.e. $\Lambda^2_{+}{\mathbb{R}}^4
                         \cap\Lambda^2_{-}{\mathbb{R}}^4=\emptyset$.
\end{proof} 
\end{proof}

\noindent
Having understood the structure of the manifold $G(2,4)$, i.e. the
parameter space of two planes in ${\mathbb{R}}^4$, from the 
decomposition into two spheres sitting in the spaces of 
self- and antiselfdual two forms, respectively, we want now discuss 
the geometries arising from restricting the parameter space of 
two planes to submanifolds of $G(2,4)$.

\noindent
\subsection{Grassmanian of complex planes $\mathcal{C}(\,{\mathbb{R}}^4\,)$}
\label{SubSec_ComPla}

\noindent
A complex structure is given by some 
$J:{\mathbb{R}}^4\rightarrow{\mathbb{R}}^4$ with $J^2=-\unity$.\\

\noindent
Each point $x_J\in S^2_{+}$ defines trough\\
\begin{eqnarray*}
       <J\xi\,,\,\eta> ~=~ <x_J\,,\,\xi\wedge\eta>  
\end{eqnarray*}\\
a complex structure and we can construct the map $J$ out of $x_J$:\\
\begin{eqnarray*}
      x_J &=& \sum\limits_{i<j} a_{ij} P_{+}\left(dx^i\wedge dx^j\right)\\
          &&\\
  <x_J\,,\,\xi\wedge\eta>\cdot vol &=& \ast x_J\wedge\xi\wedge\eta\\
                      &=& \underbrace{x_J\wedge\xi}_{\ast J\xi}\,\wedge\,\eta
\end{eqnarray*}\\
Then it is is simple exercise to compute $J$ in terms of $x_J$ which 
is given  by\\
\begin{eqnarray*}
      J &=& \left(
                \begin{array}{cccc}
                    ~0    & -a_{12} & -a_{13} & -a_{23} \\
                   a_{12} &   ~0    & -a_{23} & \hspace{1.7ex}a_{13} \\
                   a_{13} & \hspace{1.7ex}a_{23} &    ~0   & -a_{12} \\
                   a_{23} & -a_{13} & \hspace{1.7ex}a_{12} &  ~0      
                \end{array}
            \right)
\end{eqnarray*}\\
and in fact squares to the $-\unity$. Fixing a complex structure is 
equivalent to considering the subset of complex 2-planes in G(2,4) 
with respect to the complex structure chosen before. The
corresponding submanifold is simply\\   
\begin{eqnarray*}
     \mathcal{C}(\,{\mathbb{R}}^4\,) &=& \hbox{point}\times S^2_{-}~.
\end{eqnarray*}\\
This is the situation we found in example \ref{BSP1}.

\noindent
\begin{rem}
In principle one could have played the game the other way around, i.e.
fixing a point in $S^2_{-}$ instead. The difference is a change in the 
orientation. Since we usually stick to the orientation we have started 
with this possibility has to be excluded.
\end{rem}

\noindent
\subsection{Grassmanian of Lagrangian planes 
            $\mathcal{L}(\,{\mathbb{R}}^4\,)$}
\label{SubSec_LagrPla}

There exists other geometric structures. A very popular one and 
closely related to K\"ahler forms are symplectic geometries. In 
flat space this geometry is again defined via the K\"ahler form. 
We define the Grassmanian of Lagrangian 2-planes as the set on 
which the K\"ahler form restricts to zero, i.e.\\
\begin{eqnarray}\label{eq:LagrangianPlanes}
    \omega_J(\xi,\eta) &=& <J\xi\,,\,\eta> ~=~ <x_J\,,\,\xi\wedge\eta>~=~0~.
\end{eqnarray}\\
Therefore Lagrangian planes are somehow opposite to complex planes.\\
%%%%%%%%%%%%%%%%%%%%%%%%%%%%%%%%%%%%%%%%%%%%%%%%%%%%%%%%%%%%%%%%%%%%%%%%%%%
%  Etwas obskure Notation
%%%%%%%%%%%%%%%%%%%%%%%%%%%%%%%%%%%%%%%%%%%%%%%%%%%%%%%%%%%%%%%%%%%%%%%%%%%%
%1.) $\xi\wedge\eta\in S^2_{-}$
%
%\begin{eqnarray*}
%\omega_J(\xi,\eta)~=~<x_J,\xi\wedge\eta>&=&<\ast\,x_J,\ast\,\xi\wedge\eta>\\
%    \ast\,x_J\,\wedge\,\xi\wedge\eta&=& x_J\,\wedge\,\ast(\xi\wedge\eta) \\
%          x_J\,\wedge\,\xi\wedge\eta&=& -x_J\,\wedge\,\xi\wedge\eta \\
%         &&\\
%    \omega_J|_{S^2_{-}} &=& 0 
%\end{eqnarray*} 
%
%
%2.) $\xi\wedge\eta\in S^2_{+}$%
%
%\begin{eqnarray*}
%\omega_J(\xi,\eta) &=& <x_J\,,\,\xi\wedge\eta>\\
%                   &=& x_J\,\wedge\,\xi\wedge\eta \\
%                   &=& \sum\limits_{i<j} a_{ij} P_{+}(dx^i\wedge dx^j)
%                       \,\wedge\, 
%                       \sum\limits_{k<l} b_{kl} P_{+}(dx^k\wedge dx^l)\\
%                   &=& \frac{1}{2}\,<\vec{a}\,,\,\vec{b}> 
%                       ~\buildrel ! \over = 0~
%                       \hbox{  Hessische Normalform !} 
%\end{eqnarray*} 
%%%%%%%%%%%%%%%%%%%%%%%%%%%%%%%%%%%%%%%%%%%%%%%%%%%%%%%%%%%%%%%%%%%%%%%%%%

\noindent
Choose $\xi\wedge\eta~=~P_{+}(\xi\wedge\eta)\,+\,P_{-}(\xi\wedge\eta) 
                     ~=~\omega_{+}\,+\,\omega_{-}$.\\
\begin{eqnarray*}
\omega_J(\xi,\eta) &=& <x_J\,,\,\xi\wedge\eta>\\
                   &=& <x_J\,,\,\omega_{+}+\omega_{-}>\\
                   &=& <x_J\,,\,\omega_{+}>\\
                   &=& \frac{1}{2}\,<\vec{a}\,,\,\vec{b}> 
                       ~\buildrel ! \over = 0~
                       \hbox{  Hessische Normalform !} 
\end{eqnarray*}\\ 
with $\vec{a}$ a representative for $x_J$ and $\vec{b}$ a representative 
of $\omega_{+}$. Therefore all planes parametrised by $S^2_{-}$ and
those parametrised by the circle $S^1_{+}\subset S^2_{+}$ orthogonal 
to $\vec{a}$, i.e. the complex structure,  satisfy the defining constraint (\ref{eq:LagrangianPlanes}).\\
\begin{eqnarray*}
     \mathcal{L}(\,{\mathbb{R}}^4\,) &=& S^1\times S^2_{-}
\end{eqnarray*}\\
The $S^1$ part is the intersection of $S^2_{+}$ with the plane orthogonal to
$x_J\in S^2_{+}$.

\noindent
Taking into account that $\mathcal{L}(\,{\mathbb{R}}^4\,) ~=~ U(2)/SO(2)$
as a homogeneous space one can reinterpret the topology of  
$\mathcal{L}(\,{\mathbb{R}}^4\,)$ in terms of $U(2)$ as indicated below:
\begin{displaymath}\begin{CD}
                &          &       U(2)       &           &             \\
                &          &     \downarrow   &           &             \\
                &          & U(1)\times SU(2) &           &             \\
  \hbox{Phase } & \swarrow &                  & \searrow  & \hbox{ Hopf}\\
        S^1_{+} &          &                  &           & S^2_{-}     \\
\end{CD}\end{displaymath}

\noindent
Let us concentrate on the interpretation of $S^1_{+}$ as the phase of $U(2)$, i.e. we want to show, that fixing the point in $S^1_{+}$ corresponds to 
singling out all elements in $U(2)$ with the same phase.\\ 
To that purpose we choose a special complex structure, say:\\
\begin{eqnarray*}
    x_J &=& (\,1,\,0,\,0\,)
\end{eqnarray*}\\
which is\\
\begin{eqnarray*}
      J &=& \left(
                \begin{array}{cccc}
                    ~0~ & -1~ & ~0~ & ~0~ \\
                    ~1~ & ~0~ & ~0~ & ~0~ \\
                    ~0~ & ~0~ & ~0~ & -1~ \\
                    ~0~ & ~0~ & ~1~ & ~0~      
                \end{array}
            \right).
\end{eqnarray*}\\
This choice of $x_J$ corresponds to the K\"ahler 
form $\omega$ in (\ref{Kaehler}).
Using $J$ a real vector can be made into a complex one by the map\\
\begin{eqnarray}\label{ComplCoord}
    (\,a_1,\,a_2,\,a_3,\,a_4\,)^T \mapsto 
    \left(
          \begin{array}{c}
              a_1+i\cdot a_2\\
              a_3+i\cdot a_4
          \end{array}
    \right).
\end{eqnarray}\\
Each element of $U(2)$ can be represented by a $2\times 2$-matrix with 
the common additional restrictions.\\ 
\begin{eqnarray*}
    U &=& \left(
             \begin{array}{cc}
                a_1+i\cdot a_2 & b_1+i\cdot b_2 \\
                a_3+i\cdot a_4 & b_3+i\cdot b_4 
             \end{array}
          \right)
\end{eqnarray*}\\
The projection of the corresponding plane $\eta = a\wedge b$ into the 
``+''-eigenspace and parametrising the projection by the coordinates of
$S^1_{+}$ yields to\\ 
\begin{eqnarray*}
   P_{+}\left(a\wedge b\right) 
   &=& \hspace{2ex}\underbrace
        {
          \left\{\,
                a_{\left[1\right.}b_{\left.2\right]}
               +a_{\left[3\right.}b_{\left.4\right]}
          \,\right\}
        }_{0}\,\underbrace{P_{+}(dx^1\wedge dx^2)}_{\omega}\\
   &&   +\underbrace
         {
           \left\{\,
                a_{\left[1\right.}b_{\left.3\right]}
               -a_{\left[2\right.}b_{\left.4\right]}
           \,\right\}
         }_{\cos\alpha}\,\underbrace{P_{+}(dx^1\wedge dx^3)}_{\MyRe\,\Omega}\\
   &&   +\underbrace
         {
           \left\{\,
                a_{\left[2\right.}b_{\left.3\right]}
               +a_{\left[1\right.}b_{\left.4\right]}
           \,\right\}
         }_{\sin\alpha}\,\underbrace{P_{+}(dx^2\wedge dx^3)}_{\MyIm\,\Omega}
\end{eqnarray*}\\
In Fig.~\ref{figureG24a} we also show the sphere $S^2_{+}$ spanned by 
the three directions $\omega$,
$\MyRe\Omega$ and $\MyIm\Omega$, which follow from the differential 
forms defined in eq.~(\ref{Kaehler}) and eq.~(\ref{ComplVolume}), 
respectively. 
But according to (\ref{ComplCoord}) the complex coordinates are now
$z^1=x^1+x^2$ and $z^2=x^3+x^4$.\\

\noindent
One can compute the determinant of $U(2)$ which yields\\
\begin{eqnarray*}
   \det\,U &=& \underbrace
               {
                  \left\{\,
                           a_{\left[1\right.}b_{\left.3\right]}
                          -a_{\left[2\right.}b_{\left.4\right]}
                  \,\right\}
               }_{\cos\alpha}
              +\,i\, 
               \underbrace
               {
                  \left\{\,
                           a_{\left[1\right.}b_{\left.4\right]}
                          +a_{\left[2\right.}b_{\left.3\right]}
                  \,\right\}
               }_{\sin\alpha}
           ~=~ e^{i\alpha}~.
\end{eqnarray*}\\
So it is indeed the phase of the $U(2)$ element, which is parametrised 
by the $S^1_{+}$.

%%%%%%%%%%%%%%%%%%%%%%%%%%%%%%%%%%%%%%%%%%%%%%%%%%%%%%%%%%%%%%%%%%
%    FIGURE: G24.eps
%%%%%%%%%%%%%%%%%%%%%%%%%%%%%%%%%%%%%%%%%%%%%%%%%%%%%%%%%%%%%%%%%%
\parbox{\textwidth}
{
  \refstepcounter{figure}
  \label{figureG24a}
  \begin{center}
  \makebox[6cm]
  {
     \epsfxsize=6.5cm
     \epsfysize=4cm
     \epsfbox{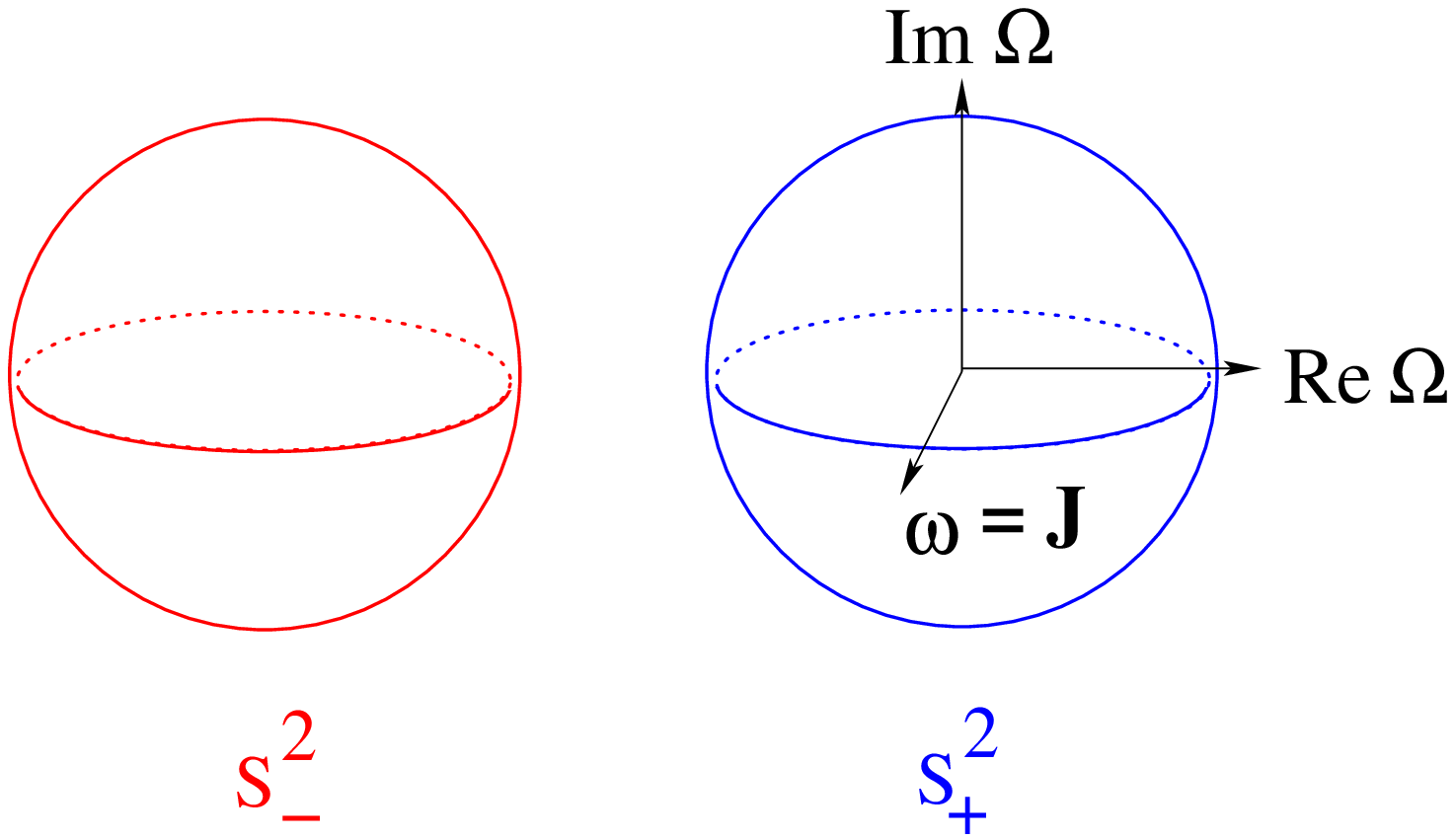}
  }
  \end{center}
  \center{{\figurefont Fig.~\thefigure:} G(2,4)}
}\\
%%%%%%%%%%%%%%%%%%%%%%%%%%%%%%%%%%%%%%%%%%%%%%%%%%%%%%%%%%%%%%%%%%
%
%%%%%%%%%%%%%%%%%%%%%%%%%%%%%%%%%%%%%%%%%%%%%%%%%%%%%%%%%%%%%%%%%%

\noindent
\subsection{Grassmanian of Special Lagrangian planes 
            $\mathcal{SL}(\,{\mathbb{R}}^4\,)$}

Combining the results from subsection \ref{SubSec_ComPla} and subsection  
\ref{SubSec_LagrPla} it becomes very simple to analyse the 
Grassmanian of Special Lagrangian planes. 
The Special Lagrangian planes are defines as those Lagrangian planes 
which share the same phase. But then obviously\\ 
\begin{eqnarray*}
       \mathcal{SL}(\,{\mathbb{R}}^4\,) ~=~ \hbox{point}\times S^2_{-}
\end{eqnarray*}
\begin{displaymath}\begin{CD}
                &          &       U(2)       &           &             \\
                &          &     \downarrow   &           &             \\
                &          & U(1)\times SU(2) &           &             \\
  \hbox{Phase } & \swarrow &                  & \searrow  & \hbox{ Hopf}\\
        S^1_{+} &          &                  &           & S^2_{-}     \\
\end{CD}\end{displaymath}\\
On the other side this point can be seen as defining a certain complex 
structure $\tilde{J}$, so that\\ 
\begin{eqnarray*}
    \mathcal{SL}_{J}(\,{\mathbb{R}}^4\,) &=&  
    \mathcal{C}_{\tilde{J}}(\,{\mathbb{R}}^4\,)~.
\end{eqnarray*}\\
It is convenient to select the point with $\alpha\,=\,0$, i.e. 
$\tilde{J}\,=\,\MyRe\,\Omega$ and the Special Lagrangian planes with 
respect to the pair ($\omega,\,\Omega$) can be understood as complex planes 
with respect to the K\"ahler structure $\tilde{J}\,=\,\MyRe\,\Omega$.\\ 

\noindent
The detailed investigation we finished here made it clear what is
meant by the notion of a $SU(2)$-calibration. The group 
$SU(2)\subset SO(4)$ can be understood as arising from 
the reduction of structure group of a general Riemannian 
2-manifold embedded in a 4-dimensional space in order to 
be consistent with the geometric structures we considered 
(complex or special Lagrangian).   In the next section we 
would like to understand the origin of such a geometrical constraint 
from the presence of supersymmetry.   

%%%%%%%%%%%%%%%%%%%%%%%%%%%%%%%%%%%%%%%%%%%%%%%%%%%%%%%%%%%%%%%%%%%%%%%%%%%
%
%%%%%%%%%%%%%%%%%%%%%%%%%%%%%%%%%%%%%%%%%%%%%%%%%%%%%%%%%%%%%%%%%%%%%%%%%%%
%\section{2-cycle equation}
%
%\begin{eqnarray*}
%   G: M&\longrightarrow& G(2,4)\\
%      x&\mapsto& T_x M\\
%     (f,g)_x&\mapsto& df_x\wedge dg_x ~=~ \sum f_{x_i}dx^i~\wedge~
%                                          \sum g_{x_j}dx^j
%                                      ~=~ \sum\limits_{i<j} f_{\left[x_i
%                                                \right.}g_{\left.x_j\right]}
%                                           dx^i\wedge dx^j
%\end{eqnarray*}
%
%
%\begin{eqnarray*}
%     P_{+}\circ G : M &\longrightarrow& \Lambda_{+}^2\\
%                    x &\mapsto& (\,0,\,1,\,0\,)
%\end{eqnarray*}
%%%%%%%%%%%%%%%%%%%%%%%%%%%%%%%%%%%%%%%%%%%%%%%%%%%%%%%%%%%%%%%%%%%%%%%%%%

\section{Linear Algebra of D4-D2 System with Flux}
\label{Sec_StringPict}

We consider again an example consisting of a system of 
a D2 and a D4-brane with additional magnetic 
flux on the D4-brane. The BPS-equations (\ref{BPS1}) and (\ref{BPS2}) 
are derived from the 
quantisation of open strings \cite{Berkooz:1996km}. 
Equation (\ref{BPS1}) contains the 
modification of the BPS-equations of a single D4-brane 
due to presence of the magnetic B-field which is taken into account 
by acting with the rotation matrix $\tilde R$ asymmetrically on the
right ($\varepsilon$) and left ($\tilde\varepsilon$) moving Killing 
spinors. Its angles $\tilde\varphi_{ij}\,=\,\tan\,F_{ij}$ corresponds to 
the flux on the D4-brane.  
The second equation (\ref{BPS2}) 
describes a general rotation $R$ of the D2-brane 
inside the D4-brane\footnote
%%%%%%%%%%%%%%%%%%%%%%%%%%%%%%%%%%%%%%%%%%%%%%%%%%%%%%%%%%%%%%
% FOOTNOTE
%%%%%%%%%%%%%%%%%%%%%%%%%%%%%%%%%%%%%%%%%%%%%%%%%%%%%%%%%%%%%%
{
  Up to now this is not an eigenvalue equation, since it 
  relates only $\varepsilon$ to $\tilde{\varepsilon}$.
}:\\
%%%%%%%%%%%%%%%%%%%%%%%%%%%%%%%%%%%%%%%%%%%%%%%%%%%%%%%%%%%%%%
%
%%%%%%%%%%%%%%%%%%%%%%%%%%%%%%%%%%%%%%%%%%%%%%%%%%%%%%%%%%%%%% 
\begin{eqnarray}\label{BPS1}
  \Gamma_{01234}\,\ROrder{\tilde{R}}\,\varepsilon
  &=&(\ROrder{\tilde{R}})^{-1}\tilde{\varepsilon}\\
  \Gamma_{012}\,\ROrder{R}\,\varepsilon
  &=& \ROrder{R}\,\tilde{\varepsilon}\label{BPS2}
\end{eqnarray}\\
Here the matrices of symmetric and asymmetric rotation are\\   
\begin{eqnarray}\label{SymmRot}
     \ROrder{R}  &=& e^{\varphi_{12}\Sigma_{12}}\cdot\ldots\cdot
                     e^{\varphi_{34}\Sigma_{34}}\\
     \ROrder{\tilde{R}}  
                 &=& e^{\tilde{\varphi}_{12}\Sigma_{12}}\cdot\ldots\cdot 
                     e^{\tilde{\varphi}_{34}\Sigma_{34}}\label{AsymmRot}
\end{eqnarray}\\
with $\Sigma_{ij}\,=\,\Gamma_{ij}/2$ the generators of $Spin(4)$.
The arrow indicates the order of the exponentials. The ordering 
in the above expressions is due to ascending tuples of numbers 
$(12)\,\leq\,(ij)\,\leq\,(34)$. 
The opposite ordering is denoted by\\   
\begin{eqnarray*}
     \LOrder{R}  &=& e^{\varphi_{34}\Sigma_{34}}\cdot\ldots\cdot 
                     e^{\varphi_{12}\Sigma_{12}}
\end{eqnarray*}\\
and for instance the following relation holds:\\
\begin{eqnarray*}
     (\ROrder{R})^{-1}  &=& \LOrder{R^{-1}}.
\end{eqnarray*}

\subsection{Choosing a gauge}
\label{SubSectionGaugeFix}

The most general flux is given by the antisymmetric tensor below:\\ 
\begin{eqnarray}\label{FluxInGeneral}
  F &=& \left(\,
                \begin{array}{cccc}
                   0 & F_{12} & F_{13} & F_{14} \\
                  -F_{12} & 0 & F_{23} & F_{24} \\
                  -F_{13} & -F_{23} & 0 & F_{34} \\ 
                  -F_{14} & -F_{24} & -F_{34} & 0
                \end{array}\,
        \right)~.
\end{eqnarray}\\
By choosing proper coordinates (for details see appendix \ref{CoordGauge}) 
it can be put into the form\\ 
\begin{eqnarray}\label{FluxNormalForm}
  \tilde{F} &=& \left(\,
                \begin{array}{cccc}
                   0 & \tilde{F}_{12} & 0 & 0 \\
                  -\tilde{F}_{12} & 0 & 0 & 0 \\
                      0   & 0 & 0 & \tilde{F}_{34} \\ 
                      0   & 0 & -\tilde{F}_{34} & 0
                \end{array}\,
                \right)~.
\end{eqnarray}\\
Here\\
\begin{eqnarray}\label{EffektiveParameter12}
   \tilde{F}_{12} &=& \frac{1}{\sqrt{2}}\,\sqrt{\sum_{i<j} F_{ij}^2 
                                                ~+~ \sqrt{\delta}}\\
   \tilde{F}_{34} &=& \frac{1}{\sqrt{2}}\,\sqrt{\sum_{i<j} F_{ij}^2 
                                                ~-~ \sqrt{\delta}}
                 \label{EffektiveParameter34}
\end{eqnarray}\\
with\\
\begin{eqnarray*}
 \delta &=& \left[\,
                    (F_{13}+F_{24})^2+(F_{12}-F_{34})^2+(F_{14}-F_{23})^2\,
            \right]\cdot
            \left[\, 
                    (F_{13}-F_{24})^2+(F_{12}+F_{34})^2+(F_{14}+F_{23})^2\,
            \right]~.
\end{eqnarray*}\\
The both factors in $\delta$ are related to the selfdual and 
antiselfdual parts of $F$.\\
\noindent
Choosing this gauge (see appendix \ref{LiftOfAction}) 
eq.~(\ref{AsymmRot}) simplifies 
since the corresponding matrix is now\\ 
\begin{eqnarray}\label{GaugedAsymmRotat}
\ROrder{\tilde{R}}  
                 &=& e^{\tilde{\varphi}_{12}\Sigma_{12}}\cdot 
                     e^{\tilde{\varphi}_{34}\Sigma_{34}}.
\end{eqnarray}

\subsubsection{The symmetric form of the BPS equations}

For practical purposes it is convenient to write the rotation matrix 
$\ROrder{R}$ in the form\\ 
\begin{eqnarray}
    \ROrder{R} &=& e^{\varphi_{12}\Sigma_{12}}\cdot
                   \underbrace{
                                e^{\varphi_{13}\Sigma_{13}}\cdot\ldots\cdot
                                e^{\varphi_{24}\Sigma_{24}}
                              }_{
                                  \ROrder{\mathfrak{R}}
                              }\cdot
                   e^{\varphi_{34}\Sigma_{34}}
\end{eqnarray}\\
with $\ROrder{\mathfrak{R}}$ containing all angles with the exception of 
$\varphi_{12}$ and $\varphi_{34}$.
Now the BPS-equations (\ref{BPS1}) and (\ref{BPS2}) can be rewritten as:\\
\begin{eqnarray*}
        \ROrder{\tilde{R}}\,\Gamma_{01234}\,\ROrder{\tilde{R}}\,\varepsilon
        &=&\tilde{\varepsilon}\\
        (\ROrder{R})^{-1}\,\Gamma_{012}\,\ROrder{R}\,\varepsilon
        &=& \tilde{\varepsilon}
\end{eqnarray*}\\
Intertwining the rotation matrices with the other gamma matrices ends up 
with:\\
\begin{eqnarray*}
        \Gamma_{01234}\,\ROrder{\tilde{R}}\,\ROrder{\tilde{R}}\,\varepsilon
        &=&\tilde{\varepsilon}\\
        \Gamma_{012}\,
        e^{-\varphi_{34}\Sigma_{34}}\,
        \LOrder{\mathfrak{R}}\,\ROrder{\mathfrak{R}}\,
        e^{ \varphi_{34}\Sigma_{34}}
        \,\varepsilon
        &=& \tilde{\varepsilon}
\end{eqnarray*}\\
Comparing the two left hand sides of the last set of equations and skipping 
the redundant $\Gamma_{012}$, the expression reads:\\ 
\begin{eqnarray*}
     \Gamma_{34}\,\LOrder{\tilde{R}}\,
                  e^{-\varphi_{34}\Sigma_{34}}\,
                  \LOrder{\mathfrak{R}}\,\ROrder{\mathfrak{R}}\,
                  e^{\varphi_{34}\Sigma_{34}}\,
                  \ROrder{\tilde{R}}\;\;(\ROrder{\tilde{R}}\,\varepsilon)
        &=& (\ROrder{\tilde{R}}\,\varepsilon)
\end{eqnarray*}\\
Multiplying both sides with $e^{\varphi_{34}\Sigma_{34}}$ and reordering 
the last expressions slightly yields\\ 
\begin{eqnarray*}
     \Gamma_{34}\,\LOrder{\tilde{R}}\,
                  \LOrder{\mathfrak{R}}\,\ROrder{\mathfrak{R}}\,
                  \ROrder{\tilde{R}}\;\;
                  (e^{\varphi_{34}\Sigma_{34}}
                   \ROrder{\tilde{R}}\,\varepsilon)
        &=& (e^{\varphi_{34}\Sigma_{34}}\ROrder{\tilde{R}}\,\varepsilon)
\end{eqnarray*}\\
Thus the problem is reduced to determine the $+1$ eigenvalues of\\
\begin{eqnarray}\label{BPSSYMM} 
    \Gamma_{34}\,\LOrder{\tilde{R}}\,\LOrder{\mathfrak{R}}\,
                 \ROrder{\mathfrak{R}}\,\ROrder{\tilde{R}}\;\;
    \eta &=& +\,\eta
\end{eqnarray}

\subsubsection{The structure of $Spin(4)$}
\label{StructureOfSpin4}

The point to note is that $\ROrder{\mathfrak{R}}\,\ROrder{\tilde{R}}$ 
represents the general element of the group $Spin(4)$. So we can use 
any representation of the general group element we want only guided 
by practical needs. Combining this freedom with the following standard 
properties of $Spin(4)$ eq.~(\ref{BPSSYMM}) can be solved completely.\\

\noindent
For this purpose we use the group isomorphism 
$Spin(4)\equiv SU(2)\times SU(2)$. The splitting reflects the splitting
of $\Lambda^2({\mathbb{R}}^4)$ with respect to the Hodge star operation
and the isomorphism mentioned above is constructed by introducing the 
following selfdual and antiselfdual combinations of generators\\
\begin{eqnarray*}
     T_1 ~=~ \frac{\Sigma_{12}\,+\,\Sigma_{34}}{2}
     &\hspace{5ex}&
     R_1 ~=~ \frac{\Sigma_{12}\,-\,\Sigma_{34}}{2}\\
     T_2 ~=~ \frac{\Sigma_{13}\,-\,\Sigma_{24}}{2}
     &\hspace{5ex}&
     R_2 ~=~ \frac{\Sigma_{13}\,+\,\Sigma_{24}}{2}\\
     T_3 ~=~ \frac{\Sigma_{23}\,+\,\Sigma_{14}}{2}
     &\hspace{5ex}&    
     R_3 ~=~ \frac{\Sigma_{23}\,-\,\Sigma_{14}}{2}
\end{eqnarray*}\\
whose algebra is\\
\begin{eqnarray*}
   [\,T_i,\,T_j\,] &=& \epsilon_{ijk}\,T_k\hspace{5ex}
   [\,R_i,\,R_j\,] ~=~ \epsilon_{ijk}\,R_k\hspace{5ex}
   [\,T_i,\,R_j\,] ~=~  0.
\end{eqnarray*}\\
Since in eq.~(\ref{BPSSYMM}) the generators $\Sigma_{12}$ and $\Sigma_{34}$
appear only in $\tilde{R}$ and the other only in $\mathfrak{R}$ we use the 
freedom to change the representation of the general element of $Spin(4)$ 
by choosing an ordering in which generators combining under (anti)selfduality 
follow each other. Then the general element of $Spin(4)$ reads\\
\begin{eqnarray*}
        \ROrder{\mathfrak{R}}\ROrder{\tilde{R}} &=& 
        e^{\alpha_{3}T_{3}}\,e^{\alpha_{2}T_{2}}\,e^{\alpha_{1}T_{1}}\,\cdot\,
        e^{\beta_{3}R_{3}}\,e^{\beta_{2}R_{2}}\,e^{\beta_{1}R_{1}}
        ~=~ \LOrder{U_T}(\alpha)\cdot \LOrder{U_R}(\beta) \\[1ex]
        \LOrder{\tilde{R}}\LOrder{\mathfrak{R}} &=& 
        e^{\beta_{1}R_{1}}\,e^{\beta_{2}R_{2}}\,e^{\beta_{3}R_{3}}\,\cdot\,
        e^{\alpha_{1}T_{1}}\,e^{\alpha_{2}T_{2}}\,e^{\alpha_{3}T_{3}}
        ~=~ \ROrder{U_R}(\beta)\cdot \ROrder{U_T}(\alpha) 
\end{eqnarray*}\\
with $\alpha$ or $\beta$ the selfdual or antiselfdual combinations of the 
$\varphi_{ij}$ and ${\tilde{\varphi}}_{ij}$ respectively\footnote
%%%%%%%%%%%%%%%%%%%%%%%%%%%%%%%%%%%%%%%%%%%%%%%%%%%%%%%%%%%%%%%%%%%%%%%%
% FOOTNOTE
%%%%%%%%%%%%%%%%%%%%%%%%%%%%%%%%%%%%%%%%%%%%%%%%%%%%%%%%%%%%%%%%%%%%%%%%
{ 
  $\alpha_1\,=\,\tilde\varphi_{12}\,+\,\tilde\varphi_{34},\,
   \alpha_2\,=\,\varphi_{13}\,-\,\varphi_{24},\,
   \alpha_3\,=\,\varphi_{23}\,+\,\varphi_{14}~$ 
  and  
  $~\beta_1\,=\,\tilde\varphi_{12}\,-\,\tilde\varphi_{34},\ldots$ 
}.\\
%%%%%%%%%%%%%%%%%%%%%%%%%%%%%%%%%%%%%%%%%%%%%%%%%%%%%%%%%%%%%%%%%%%%%%%%
Finally the product of the last two expressions becomes\\
\begin{eqnarray}
   \LOrder{\tilde{R}}\LOrder{\mathfrak{R}}
   \ROrder{\mathfrak{R}}\ROrder{\tilde{R}} 
   &=& \ROrder{U_R}(\beta)\cdot \ROrder{U_T}(\alpha)\,\cdot\,
       \LOrder{U_T}(\alpha)\cdot \LOrder{U_R}(\beta)\label{Decomposition}\\
   &=& \left(\,\ROrder{U_R}(\beta)\cdot \LOrder{U_R}(\beta)\,\right)
       \,\cdot\,
       \left(\,\ROrder{U_T}(\alpha)\cdot \LOrder{U_T}(\alpha)\,\right)
\end{eqnarray}

\subsubsection{Transformation into standard form:}

Here we note only some formulas which transform the representation 
of the Lie algebra in a form where the direct product is more obvious. 
The main purpose is to study the presentation of $\Gamma_{34}$ in this 
new basis, so that the problem of finding eigenvalues of eq.~(\ref{BPSSYMM})
can be reduced to the two independent $SU(2)$ factors.\\ 
\begin{eqnarray*}
   P &=& \left(\,\begin{array}{rrrr}
                   \frac{1}{2} & \frac{1}{\sqrt{2}} & 0 & -\,\frac{1}{2} \\ 
                   \frac{1}{2} & 0 & \frac{1}{\sqrt{2}} & \frac{1}{2}\\
                -\,\frac{i}{2} & \frac{i}{\sqrt{2}} & 0 & \frac{i}{2}\\
                   \frac{i}{2} & 0 & -\,\frac{i}{\sqrt{2}} & \frac{i}{2}
                 \end{array}\,
         \right)\;\in\;SU(4)\\
\end{eqnarray*}\\
then\\ 
\begin{eqnarray*}
   \tilde{T}_1 &=& P_1^{-1}\,T_1\,P_1 ~=~ 
                   \left(\,\begin{array}{rrrr}
                         0& 0& 0& 0\\
                         0& 0& \frac{i}{2}& 0\\
                         0& \frac{i}{2} & 0& 0\\
                         0& 0& 0& 0
                   \end{array}\,\right)\\
   \tilde{T}_2 &=& P_1^{-1}\,T_2\,P_1 ~=~   
                   \left(\,\begin{array}{rrrr}
                         0& 0& 0& 0\\
                         0& 0& \frac{1}{2} & 0\\
                         0& -\frac{1}{2}& 0& 0\\
                         0& 0& 0& 0
                   \end{array}\,\right)\\
   \tilde{T}_3 &=& P_1^{-1}\,T_3\,P_1 ~=~   
                   \left(\,\begin{array}{rrrr}
                         0& 0& 0& 0\\
                         0& -\frac{i}{2}& 0& 0\\
                         0& 0& \frac{i}{2}& 0\\
                         0& 0& 0& 0
                   \end{array}\,\right)\\
   \tilde{R}_1 &=& P_1^{-1}\,R_1\,P_1 ~=~
                   \left(\,\begin{array}{rrrr}
                         \frac{i}{2}& 0& 0& 0\\
                            0& 0& 0& 0\\
                            0& 0& 0& 0\\
                            0& 0& 0& -\frac{i}{2}
                   \end{array}\,\right)\\
  \tilde{R}_2 &=& P_1^{-1}\,R_2\,P_1 ~=~
                   \left(\,\begin{array}{rrrr}
                            0& 0& 0& -\frac{1}{2}\\
                            0& 0& 0& 0\\
                            0& 0& 0& 0\\
                           \frac{1}{2}& 0& 0& 0
                   \end{array}\,\right)
\end{eqnarray*}
\begin{eqnarray*}
   \tilde{R}_3  &=& P_1^{-1}\,R_3\,P_1 ~=~
                   \left(\,\begin{array}{rrrr}
                            0& 0& 0& -\frac{i}{2}\\
                            0& 0& 0& 0\\
                            0& 0& 0& 0\\
                            -\frac{i}{2}& 0& 0& 0
                 \end{array}\,\right)
\end{eqnarray*}\\
The matrix $\Gamma_{34}$ is transformed into\\ 
\begin{eqnarray}
   \tilde{\Gamma}_{34} &=& P^{-1}\,\Gamma_{34}\,P ~=~
                  \left(\,
                     \begin{array}{cccc}
                       -i & 0 & 0 & 0 \\
                        0 & 0 & i & 0 \\
                        0 & i & 0 & 0 \\
                        0 & 0 & 0 & i
                     \end{array}\,
                  \right) 
\end{eqnarray}

\subsection{Final step: Solution}

Now, by putting all steps together, we can discuss the existence of $+1$ 
eigenvalues for each of the two $SU(2)$s separately. With the standard 
form of the last section and the decomposition of eq.~(\ref{Decomposition}) 
equation~(\ref{BPSSYMM}) reads:\\ 
\begin{eqnarray}\label{ProblemA}
   \ROrder{U_{\tilde{T}}}(\alpha)\cdot\LOrder{U_{\tilde{T}}}(\alpha) &=&
   \left(\,
           \begin{array}{cc}
                0 & -i \\
               -i &  0 
           \end{array}\,
   \right)\\
   \ROrder{U_{\tilde{R}}}(\beta)\cdot\LOrder{U_{\tilde{R}}}(\beta) &=&
   \left(\,
           \begin{array}{cc}
               i  &  0 \\
               0  & -i 
           \end{array}\,
   \right)\label{ProblemB}
\end{eqnarray}\\
It is enough to study one of the both equations. The other problem is 
completely identical. So we concentrate on the equation (\ref{ProblemA}).
If one compares the generators $\tilde{T}_i$ with the generators 
${\mathfrak{t}_i}$ given in appendix~\ref{AllesUeberSU2} one finds the 
following map $\tilde{T}_i\,=\,{\mathfrak{t}_i}$ and it is straightforward 
to write out the equation~(\ref{ProblemA}) explicitly.\\ 

\noindent
The BPS equations are obtained as follows. When we split eq.~(\ref{BPSSYMM}) 
into the two independent conditions eq.~(\ref{ProblemA}) and 
eq.~(\ref{ProblemB}) the condition of finding {\sl eigenvectors}
is translated into an {\sl identity} on the level of matrices.  
This is due to the fact that the eigenvalues of $SU(2)$ come 
in pairs $e^{i\lambda},\,e^{-i\lambda}$ and so  the existence of 
one $+1$-eigenvalue implies the existence of two $+1$-eigenvalues.\\

\noindent
There are three sets of commuting rotations,\\        
\begin{eqnarray*}
        {\rm a)}\hspace{12ex}
            [\,\Gamma_{12},\,\Gamma_{34}\,] &=& 0~,\\[0.5ex]{}
        {\rm b)}\hspace{12ex}
            [\,\Gamma_{13},\,\Gamma_{24}\,] &=& 0~,\\[0.5ex]{}
        {\rm c)}\hspace{12ex}
            [\,\Gamma_{14},\,\Gamma_{23}\,] &=& 0~,
\end{eqnarray*}\\ 
those simultaneous in 12 and 34, those in 23 and
14 and those in 24 and 13 directions. By decomposing $\tilde{R}_1^2 R_2^2$ into
these three components we then get the conditions to preserve any supersymmetry: 
\begin{eqnarray} 
    {\rm a)}\hspace{6ex} 0 &=& F_{12} \,+\, \frac{1}{F_{34}}, 
          \nonumber\\  
    {\rm b)}\hspace{6ex}0 &=& \varphi_{23}  \,-\, \varphi_{14} \,+\, 
          \arctan (F_{23})\,-\, \arctan(F_{14}) , 
          \label{susyfluxrot2}\\  
    {\rm c)}\hspace{6ex}0 &=& \varphi_{24}  \,+\, \varphi_{13} \,+\, 
          \arctan (F_{24})\,+\, \arctan(F_{13}).
          \nonumber 
\end{eqnarray}
Each line of (\ref{susyfluxrot2}) states a condition that is capable
to define a flat supersymmetric D-brane bound state by being satisfied
globally, whereas they may only be patched together locally. 
The three rotations, symmetric or asymmetric, corresponds to three different
relative $U(1)$
rotation of the two branes. Only together they generate the most general local
$SU(2)$ deformation of the globally flat cycle. One may simplify the
conditions by choosing coordinates where one of the $U(1)$ rotations is
absorbed, such that e.g. $f^\ast\MyIm\,\Omega=0$. 
The choice of relative signs in (\ref{susyfluxrot2}) 
is arbitrary and stems from a paricular 
choice of complex structure. It relates the equation with minus signs to 
the symplectic structure $f^\ast\omega$ and the one without to $f^\ast\MyIm\,
\Omega$.\\ 
%%%%%%%%%%%%%%%%%%%%%%%%%%%%%%%%%%%%%%%%%%%%%%%%%%%%%%%%%%%%%%%%%%
%    FIGURE: Intersection.eps
%%%%%%%%%%%%%%%%%%%%%%%%%%%%%%%%%%%%%%%%%%%%%%%%%%%%%%%%%%%%%%%%%%
\parbox{\textwidth}
{
  \refstepcounter{figure}
  \label{figureS2inD4}
  \begin{center}
     \mbox
        { 
         \begin{turn}{0}%
           \epsfig{file=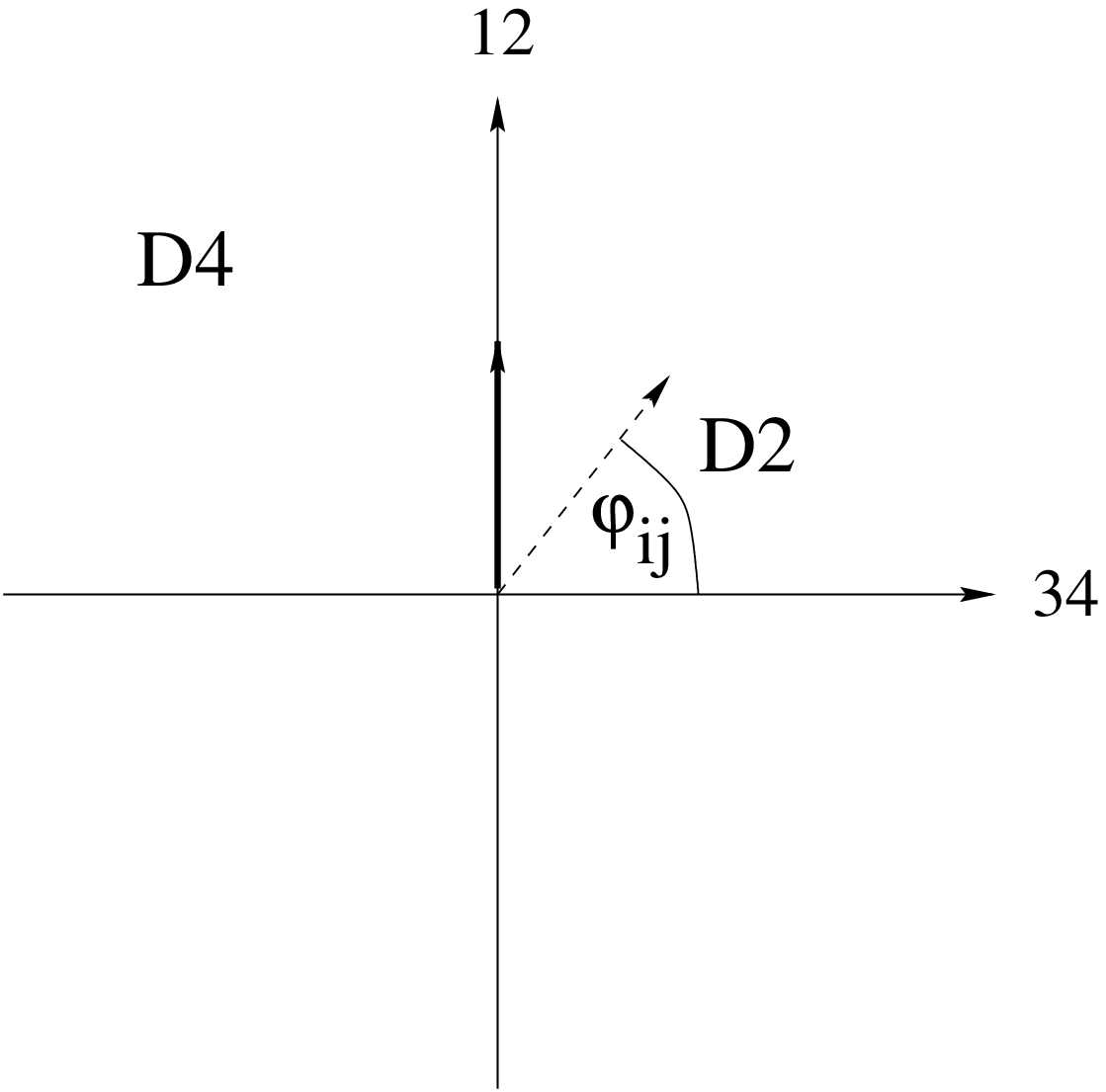,width=6cm}
         \end{turn}
        }
  \end{center}
  \center{{\figurefont Fig.~\thefigure:} Rotated D2 inside D4}
}\\[2ex]
%%%%%%%%%%%%%%%%%%%%%%%%%%%%%%%%%%%%%%%%%%%%%%%%%%%%%%%%%%%%%%%%%%
%
%%%%%%%%%%%%%%%%%%%%%%%%%%%%%%%%%%%%%%%%%%%%%%%%%%%%%%%%%%%%%%%%%%
We now follow the usual procedure to replace the flat intersecting
branes by a smooth curve $X_1(x_3,x_4)$, $X_2(x_3,x_4)$, which means 
replacing the global angles $\varphi_{ij}$ by local quantities according to 
\begin{eqnarray}
     \tan (\varphi_{ij}) ~\hat{=}~ \partial_j X_i~. 
\end{eqnarray}
Then we find 
\begin{eqnarray} \label{fomflux}
       \frac{f^\ast\omega}{f^\ast\MyRe\,\Omega}&=& -\,
       \frac{F_{23}\,-\,F_{14}}{1\,+\,F_{23}F_{14}}\, ,  \\
       \frac{f^\ast\MyIm\,\Omega}{f^\ast\MyRe\,\Omega} & =&- \, 
       \frac{F_{24}\,+\,F_{13}}{1\,-\,F_{24}F_{13}} \, .\nonumber  
\end{eqnarray}
These are the conditions that the deformed cycle preserves any 
supersymmetry in the presence of the 2-form flux on the D4-brane. 
Note, that in the case
$F_{23}=F_{14}=(*F)_{23}$ and $F_{24}=-F_{13}=(*F)_{24}$ the standard
conditions of a special Lagrangian calibration are recovered. 
Then, the field strength and the
cycle are separately supersymmetric. In (\ref{fomflux}) the deviation of the
flux from being self-dual or anti-self-dual is compensated by the deviation of
the cycle from being special Lagrangian. \\

\noindent
In this section we have seen how the supersymmetry constraints
eq.~(\ref{BPS1}) and eq.~(\ref{BPS2}) of a D2-D4 bound state  
with magnetic flux leads to the geometrical conditions 
of $SU(2)$-calibrations discussed in section \ref{Sec_G24} before.\\

\noindent
In the next section we want to discuss the derivation of the
calibration conditions for the case of $SU(d)$-calibrations 
systematically.

\section{The $SU(d)$-Cycle Equations}
\label{Sec_SLAC_D}

A d-dimensional `curve' $\Sigma^{(d)}$, embedded into $2d$-dimensional 
flat space ${\mathbb{R}}^{2d}$ with coordinates $x^i$ ($i=1,{\ldots\,},2d$) 
can be described at least locally by the zero locus of $d$ real functions 
$f^m(x^1,{\ldots\,},x^{2d})$:\\
\begin{equation}
\Sigma^{(d)} = {\mathbb{V}}(f^1,{\ldots\,},f^d) = \{(x_1,{\ldots\,},x_{2d})\,|\,
                 f^m(x^1,{\ldots\,},x^{2d})=0,\;  m=1,{\ldots\,},d\}.
                 \nonumber
\end{equation}\\
If one wants to deal with a so called supersymmetric $d$-cycle, the choice
of the functions $f^m$ is highly constrained.
To study these restrictions  we first introduce $d$ real coordinates
$\xi_i$ 
($i=1,\dots ,d$) 
which parametrise the curve $\Sigma^{(d)}$. Furthermore we consider
complex coordinates $u^i$, $u^i=x^{2i-1}+ix^{2i}$, of ${\mathbb{C}}^d$.
Then the $d$-cycle can be characterised by making the complex $u^i$ to be 
functions of the real coordinates $\xi_i$, i.e. by the following embedding 
map $i$ from $\Sigma^{(d)}$ into ${\mathbb{C}}^d$:\\
\begin{equation}
  i:\Sigma^{(d)}\longrightarrow {\mathbb{C}}^d:
    \qquad \xi_i \longrightarrow u^i(\xi_i),\quad i=1,\ldots, d.
\end{equation}\\ 
%%%%%%%%%%%%%%%%%%%%%%%%%%%%%%%%%%%%%%%%%%%%%%%%%%%%%%%%%%%%%%%%%%
%    FIGURE: Intersection.eps
%%%%%%%%%%%%%%%%%%%%%%%%%%%%%%%%%%%%%%%%%%%%%%%%%%%%%%%%%%%%%%%%%%
\parbox{\textwidth}
{
  \refstepcounter{figure}
  \label{figureIntersection}
  \begin{center}
     \mbox
        { 
         \begin{turn}{0}%
           \epsfig{file=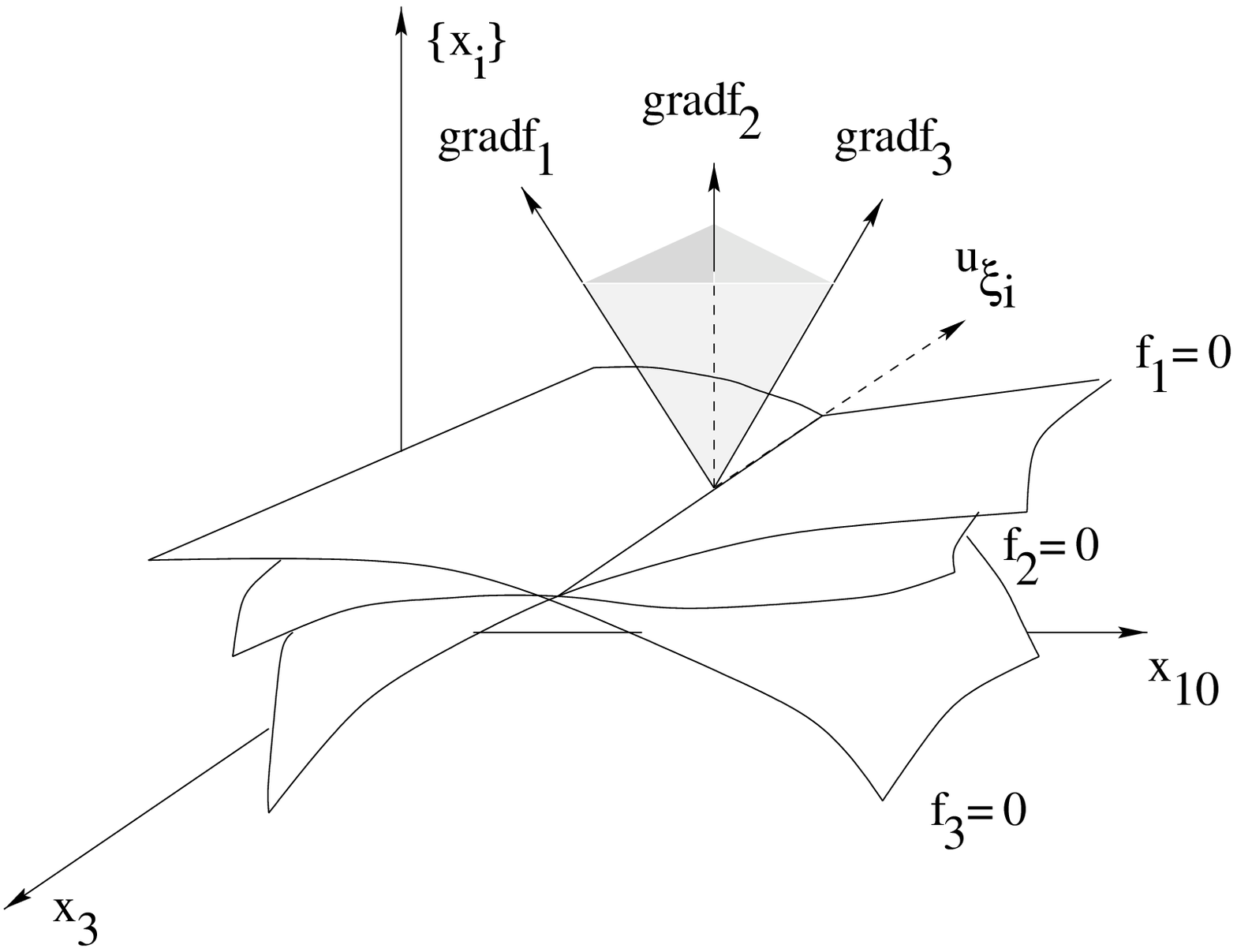,width=6cm}
         \end{turn}
        }
  \end{center}
  \center{{\figurefont Fig.~\thefigure:} The intersecting configuration}
}\\[2ex]
%%%%%%%%%%%%%%%%%%%%%%%%%%%%%%%%%%%%%%%%%%%%%%%%%%%%%%%%%%%%%%%%%%
%
%%%%%%%%%%%%%%%%%%%%%%%%%%%%%%%%%%%%%%%%%%%%%%%%%%%%%%%%%%%%%%%%%%
The intersection configuration (for the case $d=3$) is depicted in figure
\ref{figureIntersection}.\\
Now by applying the partial derivative $\partial_{\xi_k}$ to the defining 
equations $f^m$ of the $d$-cycle, we get the following relations:\\ 
\begin{equation}
  \sum_{n=1}^d (f^m_{u^n} u^n_{\xi_k} + f^m_{\bar{u}^n}
  \bar{u}^n_{\xi_k}) =0
\end{equation}\\
(here $f^m_{u^n}=\frac{\partial f^m}{\partial u^n}$, $u^n_{\xi_k}=\frac{\partial u^n}{
\partial\xi_k}$).
These can be grouped into the following matrix expressions\\
\begin{eqnarray}
\left(\begin{array}{cccc} 
        f^1_{u^1} &  f^1_{u^2} & \dots & f^1_{u^d}\\
        f^2_{u^1} &  f^2_{u^2} & \dots & f^2_{u^d} \\
        \dots &\dots &\dots &\dots \\
        f^d_{u^1} &  f^d_{u^2} & \dots &f^d_{u^d} \\
      \end{array}
\right)\cdot 
\left(\begin{array}{c}
         u^1_{\xi_k} \\
         u^2_{\xi_k} \\
         \dots \\
         u^d_{\xi_i} \\
      \end{array}
\right) &=& (-1)^{d}
\left(\begin{array}{cccc} 
        f^1_{\bar{u}^1} &  f^1_{\bar{u}^2} & \dots &f^1_{\bar{u}^d} \\
        f^2_{\bar{u}^1} &  f^2_{\bar{u}^2} & \dots &f^2_{\bar{u}^d} \\
        \dots & \dots  &\dots &\dots \\
        f^d_{\bar{u}^1} &  f^d_{\bar{u}^2} & \dots &f^d_{\bar{u}^d} \\
      \end{array}
\right)\cdot 
\left(\begin{array}{c}
         \bar{u}^1_{\xi_k} \\
         \bar{u}^2_{\xi_k} \\
         \dots \\
         \bar{u}^d_{\xi_k} \\
      \end{array}
\right)\nonumber
\end{eqnarray}\\
We will denote  the left matrix by $M$ and  the 
right matrix  as $\bar{M}$, henceforth. 
Note, the sign in front of $\bar{M}$ depends 
on the dimension $d$ of the cycle. With the help
of these matrices we can express the bared derivatives 
by the unbared ones in the following way:\\  
\begin{equation}
    \partial_k \bar{U} = (-1)^d \bar{M}^{-1}M\cdot\partial_k U 
                       = N\cdot\partial_k U .\label{lemma}
\end{equation}\\
By definition  $N$ shares the properties:\\
\begin{enumerate}
\item $N^{-1} = (-1)^d M^{-1}\bar{M} = \bar{N}$
\item $\left|\det \, N\right| = 1$
\end{enumerate}
Remembering the $d$-cycle should be supersymmetric we can ask for 
restrictions of the matrix $N$ following from this condition. It is 
well known that the notion of supersymmetric cycles \cite{Becker:1995kb} 
coincides with the notion of special Lagrangian submanifolds \cite{HL}
which can be rephrased in terms of the embedding map $i$: $\Sigma^{(d)}$ 
$\longrightarrow {\mathbb{C}}^d$ and the two conditions:\\
\begin{eqnarray}
      i^\ast \MyIm\,\Omega &=& 0  \;\;\;\;  {\rm volume\;\; minimizing}
      \nonumber\\
      i^\ast \omega &=& 0      \;\;\;\;  {\rm Lagrangian\;\; submanifold}
\label{volmin}
\end{eqnarray}\\
With $\Omega\,=\,du^1\wedge\ldots\wedge du^d$ and 
$\omega\,=\,\frac{1}{2i}\sum\limits_{i} du^i\wedge d\bar{u}^i$. 
The requirement of minimal volume reads\\
\begin{eqnarray}
 0~=~i^\ast\MyIm\,\Omega&=&\MyIm\,(du^1(\xi_1,{\ldots},\xi_d)\wedge 
                               \ldots\wedge 
                               du^d(\xi_1,{\ldots},\xi_d))\nonumber\\ 
  &=&\MyIm\,( \epsilon_{i_1{\ldots}i_d}u^{i_1}_{\xi_1}u^{i_2}_{\xi_2}{\ldots}u^{i_d}_{\xi_d})\; 
             d\xi^1\wedge{\ldots}\wedge d\xi^d\nonumber\\
\Rightarrow\;\;\;\;
0
&=&\frac{1}{2i}(\epsilon_{i_1{\ldots}i_d}u^{i_1}_{\xi_1}{\ldots}u^{i_d}_{\xi_d} -
                \epsilon_{i_1{\ldots}i_d}\bar{u}^{i_1}_{\xi_1}{\ldots}
                               \bar{u}^{i_d}_{\xi_d} )\nonumber\\
  &=&\frac{1}{2i}(\epsilon_{i_1{\ldots}i_d}u^{i_1}_{\xi_1}{\ldots}u^{i_d}_{\xi_d}
                 -\epsilon_{i_1{\ldots}i_d} N^{i_1}_{j_1}{\ldots}N^{i_d}_{j_d} u^{j_1}_{\xi_1}
                             {\ldots}u^{j_d}_{\xi_d})\nonumber\\
  &=&\frac{1}{2i}(\epsilon_{i_1{\ldots}i_d}
                 -\epsilon_{j_1{\ldots}j_d}N^{j_1}_{i_1}{\ldots}N^{j_d}_{i_d}) 
                  u^{i_1}_{\xi_1}{\ldots}u^{i_d}_{\xi_d}\nonumber\\
  &=&\frac{1}{2i}\left(1-\det\,N\right)
      \cdot\frac{\partial (u^1,{\ldots},u^d)}{\partial (\xi_1,{\ldots},\xi_d)}
     \nonumber
\end{eqnarray}\\
which  yields\\
\begin{eqnarray}
    \det \, N|_{{\mathbb{V}}(f^1,{\ldots},f^n)} = 1\;\;\;\; 
               &{\rm or\; for\; short}&\;\;\;\;
             \det\, N \equiv 1.\nonumber
\end{eqnarray}\\
For the calculation of the det-equation the following relation is useful.\\
\begin{eqnarray}
    \det\, N &\equiv& 1\;\;\;\Leftrightarrow\;\;\; 
    \det\, M - (-1)^d \det\, \bar{M}\equiv 0\nonumber
\end{eqnarray}\\
Now we turn to the second equation. 
With the canonical K\"ahler (symplectic) form $\omega$, the pull back 
operation results in\\ 
\begin{eqnarray}
  0 = i^\ast\omega &=&\frac{1}{2i}\sum\limits_{i}du^i\left(\xi_1\ldots\xi_d\right)
        \wedge d\bar{u}^i\left(\xi_1\ldots\xi_d\right)\nonumber\\
    &=& \frac{1}{2i}\sum\limits_{i}\left(\sum\limits_k u^i_{\xi_k}d\xi_k\right)
        \wedge\left(\sum\limits_l\bar{u}^i_{\xi_l}d\xi_l\right)\nonumber\\
    &=& \frac{1}{2i}\sum\limits_{k<l}\sum\limits_{i} \left[
        u^i_{\xi_k}\bar{u}^i_{\xi_l} -
        u^i_{\xi_l}\bar{u}^i_{\xi_k}   
        \right] d\xi_k\wedge d\xi_l\nonumber
\end{eqnarray}
\begin{eqnarray}
\Rightarrow\;\;\; 0 &=& \sum\limits_{i} \left[
        u^i_{\xi_k}\bar{u}^i_{\xi_l}-
        u^i_{\xi_l}\bar{u}^i_{\xi_k}   
        \right] \nonumber\\
&=& \sum\limits_{i} \left[
        u^i_{\xi_k}\left(\sum\limits_m N^i_m u^m_{\xi_l}\right)-
        u^i_{\xi_l}\left(\sum\limits_m N^i_m u^m_{\xi_k}\right)   
        \right] \nonumber\\
&=& \sum\limits_{i}
        u^i_{\xi_k}\left(\sum\limits_m N^i_m u^m_{\xi_l}\right)-
    \sum\limits_{i}
        u^i_{\xi_l}\left(\sum\limits_m N^i_m u^m_{\xi_k}\right) 
      \nonumber\\
&=& \sum\limits_{i}
        u^i_{\xi_k}\left(\sum\limits_m N^i_m u^m_{\xi_l}\right)-
    \sum\limits_{m}\left(\sum\limits_i
        u^i_{\xi_l} N^i_m\right) u^m_{\xi_k} 
      \nonumber\\
&=& \sum\limits_{i}
        u^i_{\xi_k}\left(\sum\limits_m N^i_m u^m_{\xi_l}\right)-
    \sum\limits_{i}\left(\sum\limits_m
        u^m_{\xi_l} {N^T}^i_m\right) u^i_{\xi_k} 
      \nonumber\\
&=& \sum\limits_{i,m}\left(N^i_m-{N^T}^i_m\right)u^i_{\xi_k} u^m_{\xi_l}
      \nonumber
\end{eqnarray}\\
which is  satisfied if we set $N\equiv N^T$.  
However, as it stands, this requirement is sufficient, only. Now we intend to 
give a proof that the condition is necessary, too. 

\noindent
To proof  $N\equiv N^T$ we remember some facts from
symplectic geometry especially various ways of characterising Lagrangian
planes in symplectic vector spaces. The utility of this investigation rest
on the simple observation that our conditions on the d-cycle to be a 
special Lagrangian submanifolds are in fact conditions on its tangent 
bundle, i.e. Lagrangian planes locally.\\[0.1cm]

\noindent
To begin with, we consider a complex vector space ${\mathbb{C}}^d$ furnished 
with a Hermitian structure\\
\begin{eqnarray}
   <x,y>\, = \sum\limits_i x_i\bar{y}_i = g(x,y) + i\,\sigma(x,y) \nonumber
\end{eqnarray}\\
which splits into an Euclidean metric $g$ and a symplectic form $\sigma$. 
One can check that $\sigma$ coincides with\\
\begin{eqnarray}
  \omega = \frac{1}{2i}\sum\limits_i du^i\wedge d\bar{u}^i. \nonumber
\end{eqnarray}\\
given before. Therefore we identify both 
objects. 
The two-form $\omega$ is non degenerated, antisymmetric and bilinear. 
With help of $\omega$ we can define the notion of symplectic orthogonality.
\begin{defn}
  The orthogonal complement of a vector subspace $E\in {\mathbb{C}}^d$ is
  defined by
  \begin{eqnarray}
         E^\perp  = \{ x\in {\mathbb{C}}^d \mid \omega(x,E) = 0 \} \nonumber
  \end{eqnarray}
\end{defn}
\noindent
In the special case that $E=E^\perp$ we call $E$ a Lagrangian plane. 
Obviously on a Lagrangian plane the symplectic form restricts to zero. So we
recognise the content of the constraint $i^\ast\omega=0$. It simply states 
that all tangent spaces to the supersymmetric cycle are Lagrangian planes 
embedded in the tangent space of the embedding space. Here we collect some 
facts:
\begin{enumerate}
\item $Sp(E)$ operates transitively on Lagrangian planes
\item Since $U(d)$ preserves the Hermitian form, it is contained in $Sp(E)$.
\item By $\mathcal{L}(\,{\mathbb{C}}^{d}\,)$ we denote the Gra{\ss}mannian 
      of Lagrangian planes 
\item $\lambda\in\mathcal{L}(\,{\mathbb{C}}^{d}\,)$ is characterised by 
      choosing an orthonormal basis $(a_1,\ldots ,a_n)$ with respect to 
      the Euclidean metric $g$.
      But then it is orthonormal with respect to the Hermitian form, too:\\ 
      \begin{eqnarray}
          < a_i,a_j >\, = g(a_i,a_j) + i\,\omega(a_i,a_j) 
          \buildrel !\over = \delta_{ij}, \nonumber
      \end{eqnarray}\\
      i.e. the matrix $a=(a_1,\ldots , a_n)$ is unitary. The other direction 
      works, too. 
      Hence\\ 
      \begin{eqnarray}
        \lambda\in\mathcal{L}(\,{\mathbb{C}}^{d}\,) \;\; 
        \Leftrightarrow\;\; \exists\, a\in U(d),\; 
        \lambda = a({\mathbb{R}}^d) \nonumber
      \end{eqnarray}
      %%%%%%%%%%%%%%%%%%%%%%%%%%%%%%%%%%%%%%%%%%%%%%%%%%%%%
      %   Figure LagrangianPlanes
      %%%%%%%%%%%%%%%%%%%%%%%%%%%%%%%%%%%%%%%%%%%%%%%%%%%%%
      \parbox{\textwidth}
      {
         \refstepcounter{figure}
         \label{LagrangianPlanes}
         \begin{center}
         \makebox[6cm]
         {
           \epsfxsize=6cm
           \epsfysize=6cm
           \epsfbox{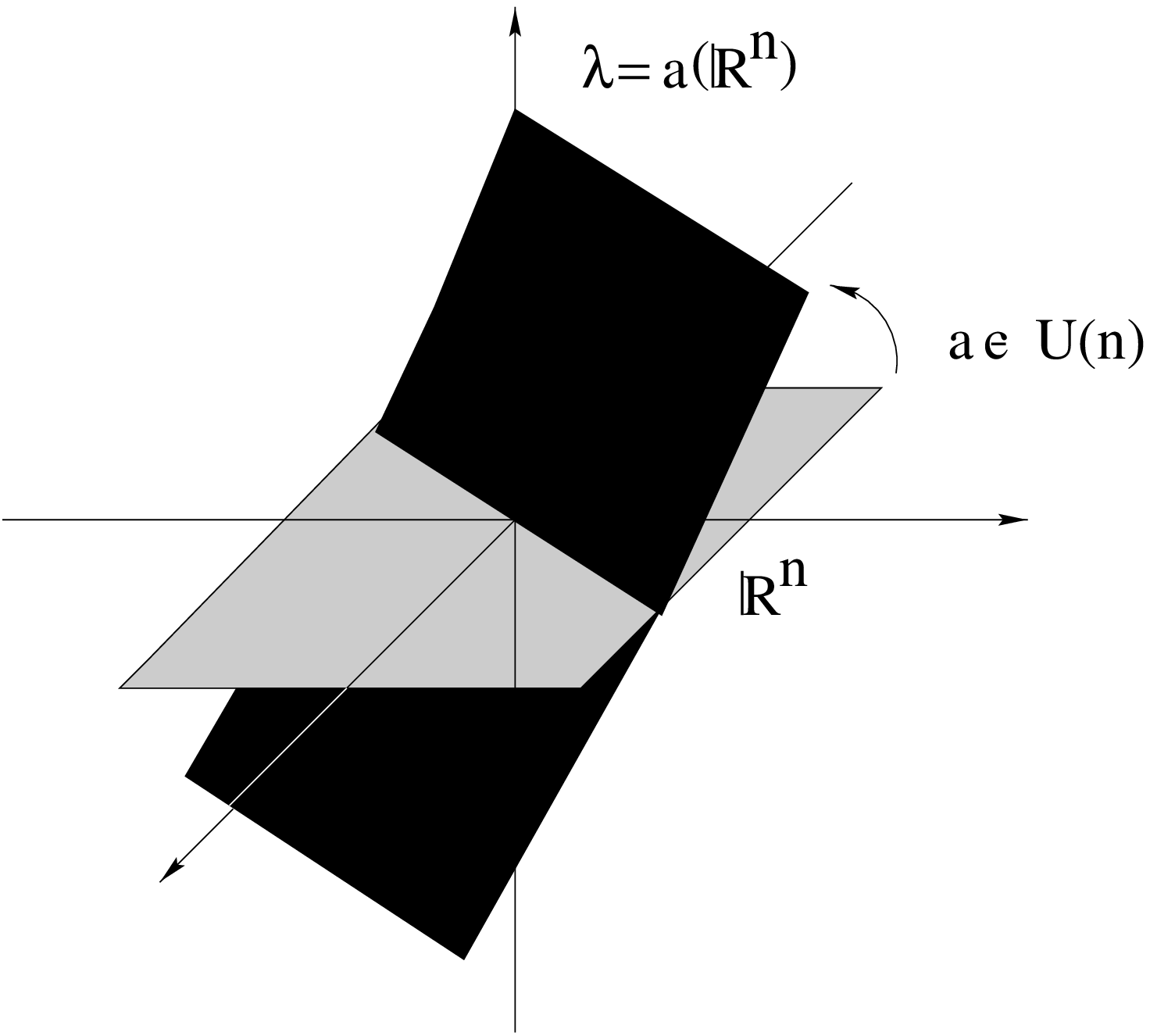}
         }
         \end{center}
         \center{{\figurefont Fig.~\thefigure:} The operation of $a$ 
                                                on Lagrangian planes}
      }
      %%%%%%%%%%%%%%%%%%%%%%%%%%%%%%%%%%%%%%%%%%%%%%%%%%%%%
      %
      %%%%%%%%%%%%%%%%%%%%%%%%%%%%%%%%%%%%%%%%%%%%%%%%%%%%%
\item Obviously each Lagrangian plane will be stabilised by any element 
      in $O(n)$, i.e. we can regard the Gra{\ss}mannian of Lagrangian planes 
      as the quotient space\\
      \begin{eqnarray}
         \mathcal{L}(\,{\mathbb{C}}^{d}\,) ~=~ \frac{U(d)}{O(d)} \nonumber
      \end{eqnarray} 
\end{enumerate}
How can we define a projection from $U(d)$ onto 
$\mathcal{L}(\,{\mathbb{C}}^{d}\,)$ ? We observe 
that two elements $a$ and $a'$ determine the same Lagrangian plane, iff\\ 
\begin{eqnarray}
 \lambda = a({\mathbb{R}}^d) ={a'}({\mathbb{R}}^d) 
             \Leftrightarrow\;\;\;
       a\bar{a}^{-1} = {a'}{\bar{a}}^{\prime -1}, \nonumber
\end{eqnarray}\\
which is constant on the $O(d)$-orbits of the fibration.
Now we can identify $\mathcal{L}(\,{\mathbb{C}}^{d}\,)$ with the image of 
the projection map\\
\begin{eqnarray}
    \pi : U(d) &\rightarrow&  \mathcal{L}(\,{\mathbb{C}}^{d}\,)\nonumber\\
           a     &\mapsto&    \lambda = a\bar{a}^{-1} \nonumber
\end{eqnarray}\\
By abuse of language we denote the matrix representative $a\bar{a}^{-1}$ of 
the Lagrangian plane $\lambda=a({\mathbb{R}}^n)$ by $\lambda$ again. 
But how can we associate the geometrical object with this artificial matrix
representative? The  connection between the matrix $\lambda$ on the one side and the concrete Lagrangian plane $\lambda$ on the other side is given through the central 
equation\\
\begin{eqnarray}
   x\in\lambda\;   \Leftrightarrow\;  x=\lambda\bar{x} \nonumber  
\end{eqnarray}\\
In the last formula we recognise the familiar equation  
(\ref{lemma}).
But now we know, that we can represent $\lambda$ as $\lambda=a\bar{a}^{-1}$
and this yields straight forward\\
\begin{eqnarray}
   \lambda^{+} &=& {\bar{a}^{-1^{+}}}a^{+} = {a^{-1}}^T a^{-1} = \bar{a}a^{-1}
                = \bar{\lambda} \nonumber\\
\Rightarrow\;\; \lambda^T &=& \lambda \nonumber
\end{eqnarray}\\
But then we can finally conclude by identifying  $\lambda = N^{-1}$ and 
performing some mild manipulations that\\
\begin{eqnarray}
      N \equiv N^T\nonumber
\end{eqnarray}
\rightline{\qed}

\noindent
In summary, all what we have done so far can be formulated 
in a short but important 
proposition which is the starting point for all further 
computations:\\[10pt]
{\bf Proposition: }{\it A $d$-cycle, represented as an intersection of 
             $d$ real  valued functions is\linebreak supersymmetric, 
             iff 
             $N\equiv N^{T}$ and $\det \, N\equiv 1$.}\\[10pt]
It will turn out to be very useful to reformulate the last proposition
$N\equiv N^{T}$ in a different, but equivalent way.
Namely, it is not difficult to show that the requirement $N\equiv N^T$ is 
equivalent to the condition that the matrix $MM^+$ should be real modulo 
$I({\mathbb{V}})$.
To prepare this reformulation we remark that by the split of the coordinates 
of ${\mathbb{R}}^{2d}$ into the coordinates of ${\mathbb{C}}^d$ they inherit an 
intrinsic meaning as the spatial and momentum variables of symplectic
geometry. This is given by\\ 
\begin{equation}
    u^i=q^i+ip^i,
\end{equation}\\
i.e. the real part of $u^i$ gets the meaning of a spatial coordinate whereas 
the $p^i$ is a momentum variable. Then we are free to define the convenient 
Poisson brackets of phase-space functions $\{ f,g\}$. This is done  in 
the standard way as\\
\begin{equation}
\{ f,g\} =  \sum_{i=1}^d
            \left(
               \frac{\partial f}{\partial q^i}\frac{\partial g}{\partial p^i}
              -\frac{\partial f}{\partial p^i}\frac{\partial g}{\partial q^i}
            \right) 
         = \sum_{i=1}^d (f_{2i-1}g_{2i}-f_{2i}g_{2i-1}),
\end{equation}\\
where $f_{2i-1}=\frac{\partial f}{
\partial q^i}=\frac{\partial f}{\partial x^{2i-1}}$ and
$f_{2i}=\frac{\partial f}{\partial p^i}=\frac{\partial f}{\partial x^{2i}}$.
Then the matrix
$MM^+$ reads\\
\begin{eqnarray}
 (MM^+)_{mn} =\,<\nabla\,f^m,\nabla\,f^n\,>\pm\, i\cdot\{ f^m,f^n\} .
\end{eqnarray}\\
So $MM^+$ is a real matrix modulo $I({\mathbb{V}})$, i.e. $N\equiv N^T$, if all 
Poisson brackets among the defining functions $f^m$ and  $f^n$ vanish:\\
\begin{equation}
\{ f^m,f^n\}\equiv 0.\label{poisson}
\end{equation}\\
So we get a more suitable set of equations for concrete calculations.\\[10pt]
{\bf Corollary: }{\it A $d$-cycle, represented as an intersection of 
             $d$ real  valued functions is\linebreak supersymmetric, 
             iff 
             $\{f^i,f^j\}\equiv 0$ and $\det \, 
             N\equiv 1$}

\subsection*{Acknowledgements:} A.M. would like to thank V.~Braun, 
A.~Karch, B.~K\"ors, D.~L\"ust and I.~Schnakenburg for discussions. 
The work of A.M. was financially supported by the Deutsche 
Forschungsgemeinschaft.

\begin{appendix}

\section{Determinant of a two dimensional metric}
\label{Sec_Det2dMetrik}

The determinant of the induced metric given in 
eq.~(\ref{Eq_induced_metric}) is
\begin{eqnarray}\label{Det_induced_metric}
|f^{\ast}g| &=& \left[
                   1+\left(\frac{\partial X_1}{\partial \xi_1}\right)^2
                    +\left(\frac{\partial X_2}{\partial \xi_1}\right)^2
            \right]
            \left[
                   1+\left(\frac{\partial X_1}{\partial \xi_2}\right)^2
                    +\left(\frac{\partial X_2}{\partial \xi_2}\right)^2
            \right]
            -
            \left[
                   \frac{\partial X_1}{\partial \xi_1}
                   \frac{\partial X_1}{\partial \xi_2}
                   +\frac{\partial X_2}{\partial \xi_1}
                   \frac{\partial X_2}{\partial \xi_2}
            \right]^2\nonumber\\
        &=&1+\left(\frac{\partial X_1}{\partial \xi_1}\right)^2
            +\left(\frac{\partial X_2}{\partial \xi_1}\right)^2
            +\left(\frac{\partial X_1}{\partial \xi_2}\right)^2
            +\left(\frac{\partial X_2}{\partial \xi_2}\right)^2\nonumber\\
        &&  +\left[
                   \left(\frac{\partial X_1}{\partial \xi_1}\right)^2
                   +\left(\frac{\partial X_2}{\partial \xi_1}\right)^2
            \right]
            \left[
                   \left(\frac{\partial X_1}{\partial \xi_2}\right)^2
                   +\left(\frac{\partial X_2}{\partial \xi_2}\right)^2
            \right]
            -
            \left[
                   \frac{\partial X_1}{\partial \xi_1}
                   \frac{\partial X_1}{\partial \xi_2}
                   +\frac{\partial X_2}{\partial \xi_1}
                   \frac{\partial X_2}{\partial \xi_2}
            \right]^2\nonumber\\
        &=&  1+\left(\frac{\partial X_1}{\partial \xi_1}\right)^2
             +\left(\frac{\partial X_2}{\partial \xi_1}\right)^2
             +\left(\frac{\partial X_1}{\partial \xi_2}\right)^2
             +\left(\frac{\partial X_2}{\partial \xi_2}\right)^2
             +\left(
                     \frac{\partial X_1}{\partial \xi_1}
                     \frac{\partial X_2}{\partial \xi_2}
                    -\frac{\partial X_2}{\partial \xi_1}
                     \frac{\partial X_1}{\partial \xi_2}
              \right)^2\nonumber\\
        &=&  \left(
                    \frac{\partial X_1}{\partial \xi_2}
                   -\frac{\partial X_2}{\partial \xi_1}
             \right)^2
            +\left(
                    \frac{\partial X_1}{\partial \xi_1}
                   +\frac{\partial X_2}{\partial \xi_2}
             \right)^2
            +\left(
                     1-
                     \frac{\partial X_1}{\partial \xi_1}
                     \frac{\partial X_2}{\partial \xi_2}
                    +\frac{\partial X_2}{\partial \xi_1}
                     \frac{\partial X_1}{\partial \xi_2}
             \right)^2\nonumber\\[1ex]
        &=& \left[f^\ast\omega\right]^2
            +\left[f^\ast\MyIm\,\Omega\right]^2
            +\left[f^\ast\MyRe\,\Omega\right]^2
\end{eqnarray}

%%%%%%%%%%%%%%%%%%%%%%%%%%%%%%%%%%%%%%%%%%%%%%%%%%%%%%%%%%%%%
% Appendix: Choice of Coordinates
%%%%%%%%%%%%%%%%%%%%%%%%%%%%%%%%%%%%%%%%%%%%%%%%%%%%%%%%%%%%%

\section{Decomposition of the adjoint representation of $Spin(4)$}
\label{CoordGauge}

The change from eq.~(\ref{FluxInGeneral}) to the normal form of 
eq.~(\ref{FluxNormalForm}) can be described quite explicitly by  
using the isomorphism of $Spin(4)$ to $SU(2)\times SU(2)$ 
constructed in subsection \ref{StructureOfSpin4} and the relation 
to the corresponding $SO$-groups as shown in the diagram below:  
\begin{eqnarray}\label{Diagramm-Spin4-SO4}
\begin{CD}
       Spin(4)     @=    SU(2)\times SU(2)\\
     @V{\pi}VV             @VV{\pi}V\\
        SO(4)      @=    SO(3)\times SO(3)   
\end{CD}
\end{eqnarray}
The two $SU(2)$ act via the adjoint representations ($SO(3)$) on the 
two 3-dimensional subspaces in the orthogonal decomposition of  
$\Lambda^2({\mathbb{R}}^4)\,=\,\Lambda^2_{+}({\mathbb{R}}^4)\oplus
                               \Lambda^2_{-}({\mathbb{R}}^4)$.
Since $F_{\mu\nu} \in \Lambda^2({\mathbb{R}}^4)$ 
we can obtain the normal form of eq.~(\ref{FluxInGeneral}) by 
computing the selfdual and antiselfdual parts of $F_{\mu\nu}$ due to  
\begin{eqnarray*}
        F  &=&  \frac{1}{2}\,(\,1\,+\,\ast\,)\,F ~+~
                \frac{1}{2}\,(\,1\,-\,\ast\,)\,F\\
    (F_{12},\ldots,F_{34}) &\mapsto& \left(\vbox{\vspace{2ex}}\right.
                                        [F_{12}+F_{34}]/2,\ldots\,
                                     \left.\vbox{\vspace{2ex}}\right)
                                     \oplus
                                     \left(\vbox{\vspace{2ex}}\right.
                                        [F_{12}-F_{34}]/2,\ldots\,
                                     \left.\vbox{\vspace{2ex}}\right)
\end{eqnarray*}
and determine the two rotations $S_1\times S_2 \in SO(3)\times SO(3)$, 
which rotate each of the two 3-vectors on the right hand side of the 
above decomposition to the standard position 
\begin{eqnarray*}
    \left(\vbox{\vspace{2ex}}\right.
         [\tilde{F}_{12}+\tilde{F}_{34}]/2,0,0\, 
    \left.\vbox{\vspace{2ex}}\right)
    \oplus
    \left(\vbox{\vspace{2ex}}\right.
         [\tilde{F}_{12}-\tilde{F}_{34}]/2,0,0\,
    \left.\vbox{\vspace{2ex}}\right).
\end{eqnarray*} 
Since the rotations preserve the norm it is quite easy to work out from 
this requirement the form of the expressions in 
eq.~(\ref{EffektiveParameter12}) and 
eq.~(\ref{EffektiveParameter34}). 
Explicit matrices $S_1$ and $S_2$ can be written down using the 
formula~(\ref{SO3Action}) in appendix \ref{AllesUeberSO3}. 
In the following we will need $S_1$ only, so we include the result for 
convenience. 
Note that the angles shown below are not unique due to the remaining 
rotation of a three vector around its own axis: 
\begin{eqnarray*}
        \theta_3  ~=~ -\,\arctan\frac{b}{a}\hspace{4ex}
        \theta_2  ~=~ \arctan \frac{c}{\sqrt{a^2+b^2}}\hspace{4ex}
        \theta_1  ~=~ 0
\end{eqnarray*}
Here $a$, $b$ and $c$ denote the selfdual components of $F_{\mu\nu}$.

%%%%%%%%%%%%%%%%%%%%%%%%%%%%%%%%%%%%%%%%%%%%%%%%%%%%%%%%%%%%%
% Appendix: Flux / Angles
%%%%%%%%%%%%%%%%%%%%%%%%%%%%%%%%%%%%%%%%%%%%%%%%%%%%%%%%%%%%%

\section{The invariant relation of flux and angles}
\label{LiftOfAction}

Eq.~(\ref{GaugedAsymmRotat}) can be read as a contraction of the 
antisymmetric tensor $\tilde\varphi_{ij}\,=\,\arctan\,\tilde{F}_{ij}$
with the antisymmetric gamma matrices $\Gamma_{ij}$.\\ 
In appendix \ref{CoordGauge} we constructed the matrix 
$S_1\times S_2\in SO(3)\times SO(3)$, which transforms a 
general $F$ into normal form $\tilde{F}$. This matrix is in 
the {\bf 6} representation of $SO(4)$. Here we need the 
corresponding element in  the {\bf 4} representation of $SO(4)$, which 
generates the same action as $S_1\times S_2$. Such an $S\in SO(4)$ 
acts like
\begin{eqnarray*}
    \tilde{F}_{ij} &=& S_{i}{}^k\,S_j{}^l\,F_{kl} ~=~ (S\,F\,S^T)_{ij}
\end{eqnarray*}
and can be constructed by going from the lower right to the lower left 
corner in diagram (\ref{Diagramm-Spin4-SO4}).
Utilising this $S\in SO(4)$ the exponent in eq.~(\ref{GaugedAsymmRotat}) 
can be written as 
\begin{eqnarray*}
   {\tilde\varphi}_{ij}\Gamma^{ij} 
                &=& S_i{}^{k}\,(\arctan\,F)_{kl}\,S_j{}^l\;
                    \Gamma^{ij}\\
                &=& (\arctan\,F)_{kl}\;
                    S_i{}^{k}\,\Gamma^{ij}\,S_j{}^l\\
                &=& (\arctan\,F)_{kl}\;
                    (S^T)^{k}{}_i\,(S^T)^l{}_j\,\Gamma^{ij}
\end{eqnarray*}
But the action of $SO(4)$ on $\Gamma^{ij}$ in the last line can be seen 
as the adjoint action of $Spin(4)$ on its Lie algebra 
$\Sigma_{ij}\,=\,\Gamma_{ij}/2$.
So we can rewrite the last transformation in terms of a $Spin(4)$ 
rotation and finally we obtain:
\begin{eqnarray*}
   e^{(\arctan\,F)_{kl}\;(S^T\,\Sigma\,S)^{kl}}
   &=&
   \pi^{-1}(S^{T})\,e^{(\arctan\,F)_{kl}\Sigma^{kl}}\,\pi^{-1}({S^T}^{-1}) 
\end{eqnarray*}

%%%%%%%%%%%%%%%%%%%%%%%%%%%%%%%%%%%%%%%%%%%%%%%%%%%%%%%%%%%%%
% Appendix: SU(2)
%%%%%%%%%%%%%%%%%%%%%%%%%%%%%%%%%%%%%%%%%%%%%%%%%%%%%%%%%%%%%

\section{$SU(2)$}
\label{AllesUeberSU2}

The generators of the Lie algebra $\mathfrak{su}(2)$ are 
\begin{eqnarray*}
   {\mathfrak{t}}_1 
       ~=~  \left(\, 
                    \begin{array}{cc}
                      0 & \frac{i}{2} \\ [2ex]
                      \frac {i}{2} & 0
                    \end{array}\,
            \right)\hspace{5ex} 
   {\mathfrak{t}}_2
       ~=~  \left(\,
                   \begin{array}{cc}
                      0 & \frac{1}{2}  \\ [2ex]
                      -\frac{1}{2}  & 0
                   \end{array}\,
            \right)\hspace{5ex} 
   {\mathfrak{t}}_3
       ~=~  \left(\, 
                   \begin{array}{cc}
                      -\frac{i}{2} & 0 \\ [2ex] 
                       0 & \frac{i}{2}
                   \end{array}\,
            \right) 
\end{eqnarray*}
The three one parameter subgroups generated by 
${\mathfrak{t}}_1\ldots {\mathfrak{t}}_3$ are
\begin{eqnarray*}
   g_1 &=& \left(\,
                   \begin{array}{cc}
                       \cos(\frac{\theta_1}{2}) & i\,\sin(\frac{\theta_1}{2})\\
                    i\,\sin(\frac{\theta_1}{2}) & \cos(\frac{\theta_1}{2})
                   \end{array}\,
           \right)\\
   g_2 &=& \left(\,\begin{array}{cc}
                       \cos(\frac{\theta_2}{2}) & \sin(\frac{\theta_2}{2})\\
                      -\sin(\frac{\theta_2}{2}) & \cos(\frac{\theta_2}{2})
                   \end{array}\,
           \right)\\
   g_3 &=& \left(\,\begin{array}{cc}
                      \exp(-i\frac{\theta_3}{2}) & 0 \\ 
                       0 & \exp(i\frac{\theta_3}{2})
                   \end{array}\,
           \right)
\end{eqnarray*}
The product of each of the three gives
{\small
\begin{eqnarray*}
 g_1\cdot g_2\cdot g_3 &=& \left(\,
 \begin{array}{rr}
      \left[
            \cos(\frac{\theta_1}{2})\,\cos(\frac{\theta_2}{2})
        -i\,\sin(\frac{\theta_1}{2})\,\sin(\frac{\theta_2}{2})
      \right]\,
      e^{-i\frac{\theta_3}{2}} & 
      \left[
            \cos(\frac{\theta_1}{2})\,\sin(\frac{\theta_2}{2}) 
        +i\,\sin(\frac{\theta_1}{2})\,\cos(\frac{\theta_2}{2})
      \right]\,
      e^{i\frac{\theta_3}{2}}\\
  -\,\left[
            \cos(\frac{\theta_1}{2})\,\sin(\frac{\theta_2}{2})
        -i\,\sin(\frac{\theta_1}{2})\,\cos(\frac{\theta_2}{2})
     \right]\,
      e^{-i\frac{\theta_3}{2}} &
     \left[
            \cos(\frac{\theta_1}{2})\,\cos(\frac{\theta_2}{2})
        +i\,\sin(\frac{\theta_1}{2})\,\sin(\frac{\theta_2}{2})
     \right]\,
      e^{i\frac{\theta_3}{2}}
 \end{array}\,\right)
\end{eqnarray*}
}\\[-1ex]

%%%%%%%%%%%%%%%%%%%%%%%%%%%%%%%%%%%%%%%%%%%%%%%%%%%%%%%%%%%%%
% Appendix: SO(3)
%%%%%%%%%%%%%%%%%%%%%%%%%%%%%%%%%%%%%%%%%%%%%%%%%%%%%%%%%%%%%

\section{$SO(3)$}
\label{AllesUeberSO3}

The $SO(3)$ action associated to the $SU(2)$ of appendix \ref{AllesUeberSU2}
is defined by the action of $SU(2)$ on the Lie algebra ${\mathfrak{su}}(2)$, 
i.e.
\begin{eqnarray*}
   \pi(U)\;{\vec{\mathfrak{t}}} &=& U\;\vec{\mathfrak{t}}\;U^{-1}
\end{eqnarray*}
with ${\vec{\mathfrak{t}}}\,=\,x^i\,{\mathfrak{t}}_i$.
The adjoint action on each generator can be computed easily and reads:
\begin{eqnarray*}
   \pi(g_1)\;{\mathfrak{t}}_1 &=& \hspace{28.5ex} 
       ~=~ \phantom{-\,}{\mathfrak{t}}_1 \\
   \pi(g_1)\;{\mathfrak{t}}_2
       &=& \left(\begin{array}{rr}
            -\,\frac{i}{2}\,\sin\theta_1 & \phantom{-\,}
               \frac{1}{2}\,\cos\theta_1 \\ 
            -\,\frac{1}{2}\,\cos\theta_1 & \frac{i}{2}\,\sin\theta_1 
           \end{array}\right)    
       ~=~  \phantom{-\,}
            \cos\theta_1 \cdot {\mathfrak{t}}_2 ~+~
            \sin\theta_1 \cdot {\mathfrak{t}}_3\\ 
    \pi(g_1)\;{\mathfrak{t}}_3
       &=& \left(\begin{array}{rr}
            -\,\frac{i}{2}\,\cos\theta_1 & -\,\frac{1}{2}\,\sin\theta_1 \\ 
               \frac{1}{2}\,\sin\theta_1 &    \frac{i}{2}\,\cos\theta_1
           \end{array}\right)\,    
       ~=~ -\,\sin\theta_1 \cdot {\mathfrak{t}}_2 ~+~ 
               \cos\theta_1 \cdot {\mathfrak{t}}_3  
\end{eqnarray*}
\begin{eqnarray*}
\pi(g_2)\;{\mathfrak{t}}_1 
          &=& \left(\,\begin{array}{rrr}
                 \;\frac{i}{2}\,\sin\theta_2 & 
                 \frac{i}{2}\,\cos\theta_2\\ 
                 \frac{i}{2}\,\cos\theta_2 & 
                 -\,\frac{i}{2}\,\sin\theta_2
              \end{array}\,\right)
          ~=~ -\,\sin\theta_2\cdot {\mathfrak{t}}_3 ~+~
                 \cos\theta_2\cdot {\mathfrak{t}}_1\\
\pi(g_2)\;{\mathfrak{t}}_2 &=& \hspace{27ex}\, 
          ~=~ \phantom{-\,}{\mathfrak{t}}_2\\
\pi(g_2)\;{\mathfrak{t}}_3 
          &=& \left(\,\begin{array}{rrr}
                 -\frac{i}{2}\,\cos\theta_2 & \frac{i}{2}\,\sin\theta_2\\ 
                 \frac{i}{2}\,\sin\theta_2 & \frac{i}{2}\,\cos\theta_2
              \end{array}\,\right)\;
          ~=~ \phantom{-\,}
              \cos\theta_2\cdot {\mathfrak{t}}_3 ~+~ 
              \sin\theta_2\cdot {\mathfrak{t}}_1
\end{eqnarray*}
\begin{eqnarray*}
\pi(g_3)\;{\mathfrak{t}}_1 
          &=& \left(\,\begin{array}{cc}
                  0 & \frac{i}{2}\,e^{-i\theta_3}\\
                  \frac{i}{2}\,e^{i\theta_3} & 0
              \end{array}\right)\hspace{6ex}
          ~=~ \phantom{-\,}
              \cos\theta_3\cdot {\mathfrak{t}}_1 ~+~
              \sin\theta_3\cdot {\mathfrak{t}}_2\\
\pi(g_3)\;{\mathfrak{t}}_2 
          &=& \left(\,\begin{array}{rr}
                 0 & \frac{1}{2}\,e^{-i\theta_3}\\
                -\frac{1}{2}\,e^{i\theta_3} & 0 
              \end{array}\,\right)\hspace{4ex}
          ~=~ -\,\sin\theta_3\cdot {\mathfrak{t}}_1 ~+~ 
                 \cos\theta_3\cdot {\mathfrak{t}}_2\\
\pi(g_3)\;{\mathfrak{t}}_3 &=& \hspace{27.5ex} 
          ~=~ \phantom{-\,}{\mathfrak{t}}_3
\end{eqnarray*}
The action rewritten with respect to the coordinates is then:
\begin{eqnarray}\label{SO3Action}
   \pi(g_1\cdot g_2\cdot g_3)\;\vec{\mathfrak{t}} &=&
       \left(\,\begin{array}{ccc}
          1 & 0 & 0 \\
          0 & \cos\theta_1 & -\sin\theta_1 \\ 
          0 & \sin\theta_1 & \cos\theta_1
       \end{array}\,\right)\cdot
       \left(\,\begin{array}{ccc}
          \cos\theta_2 & 0 & \sin\theta_2 \\
          0 & 1 & 0 \\ 
          -\sin\theta_2 & 0 & \cos\theta_2
       \end{array}\,\right)\cdot\\
       &&\hspace{27.5ex}
       \left(\,\begin{array}{ccc}
          \cos\theta_3 & -\sin\theta_3 & 0 \\
          \sin\theta_3 & \cos\theta_3 & 0 \\ 
          0 & 0 & 1
       \end{array}\,\right)\,
       \left(\begin{array}{c} x^1\\x^2\\x^3 \end{array}\right)
       \nonumber
\end{eqnarray} 

\end{appendix}

\begingroup\raggedright\endgroup

%% \CharacterTable
%%  {Upper-case    \A\B\C\D\E\F\G\H\I\J\K\L\M\N\O\P\Q\R\S\T\U\V\W\X\Y\Z
%%   Lower-case    \a\b\c\d\e\f\g\h\i\j\k\l\m\n\o\p\q\r\s\t\u\v\w\x\y\z
%%   Digits        \0\1\2\3\4\5\6\7\8\9
%%   Exclamation   \!     Double quote  \"     Hash (number) \#
%%   Dollar        \$     Percent       \%     Ampersand     \&
%%   Acute accent  \'     Left paren    \(     Right paren   \)
%%   Asterisk      \*     Plus          \+     Comma         \,
%%   Minus         \-     Point         \.     Solidus       \/
%%   Colon         \:     Semicolon     \;     Less than     \<
%%   Equals        \=     Greater than  \>     Question mark \?
%%   Commercial at \@     Left bracket  \[     Backslash     \\
%%   Right bracket \]     Circumflex    \^     Underscore    \_
%%   Grave accent  \`     Left brace    \{     Vertical bar  \|
%%   Right brace   \}     Tilde         \~}

\begin{thebibliography}{10}

\bibitem{HL}
R.~Harvey and H.~B. Lawson, ``Calibrated {G}eometries,'' {\em Acta Mathematica}
  {\bf 148} (1982) 47--157.

\bibitem{Becker:1995kb}
K.~Becker, M.~Becker, and A.~Strominger, ``Five-branes, membranes and
  nonperturbative string theory,'' {\em Nucl. Phys.} {\bf B456} (1995)
  130--152, \href{http://xxx.lanl.gov/abs/hep-th/9507158}{{\tt
  hep-th/9507158}}.

\bibitem{Witten:1997sc}
E.~Witten, ``Solutions of four-dimensional field theories via {M}-theory,''
  {\em Nucl. Phys.} {\bf B500} (1997) 3,
  \href{http://xxx.lanl.gov/abs/hep-th/9703166}{{\tt hep-th/9703166}}.

\bibitem{Smith}
A.~Fayyazuddin and D.~J. Smith, ``Localized intersections of {M}5-branes and
  four-dimensional superconformal field theories,'' {\em JHEP} {\bf 04} (1999)
  030, \href{http://xxx.lanl.gov/abs/hep-th/9902210}{{\tt hep-th/9902210}}.

\bibitem{Gauntlett:1998vk}
J.~P. Gauntlett, N.~D. Lambert, and P.~C. West, ``Branes and calibrated
  geometries,'' {\em Commun. Math. Phys.} {\bf 202} (1999) 571,
  \href{http://xxx.lanl.gov/abs/hep-th/9803216}{{\tt hep-th/9803216}}.

\bibitem{Gauntlett:1998wb}
J.~P. Gauntlett, N.~D. Lambert, and P.~C. West, ``Supersymmetric five-brane
  solitons,'' {\em Adv. Theor. Math. Phys.} {\bf 3} (1999) 91,
  \href{http://xxx.lanl.gov/abs/hep-th/9811024}{{\tt hep-th/9811024}}.

\bibitem{Barwald:1999ux}
O.~B{\"a}rwald, N.~D. Lambert, and P.~C. West, ``A calibration bound for the
  {M}-theory fivebrane,'' {\em Phys. Lett.} {\bf B463} (1999) 33,
  \href{http://xxx.lanl.gov/abs/hep-th/9907170}{{\tt hep-th/9907170}}.

\bibitem{Lust:1999pq}
D.~L{\"u}st and A.~Miemiec, ``Supersymmetric {M}5-branes with {H}-field,'' {\em
  Phys. Lett.} {\bf B476} (2000) 395--401,
  \href{http://xxx.lanl.gov/abs/hep-th/9912065}{{\tt hep-th/9912065}}.

\bibitem{Acharya:1998yv}
B.~S. Acharya, J.~M. Figueroa-O'Farrill, and B.~J. Spence, ``Planes, branes and
  automorphisms. {I}: {S}tatic branes,'' {\em JHEP} {\bf 07} (1998) 004,
  \href{http://xxx.lanl.gov/abs/hep-th/9805073}{{\tt hep-th/9805073}}.

\bibitem{Acharya:1998en}
B.~S. Acharya, J.~M. Figueroa-O'Farrill, and B.~J. Spence, ``Branes at angles
  and calibrated geometry,'' {\em JHEP} {\bf 04} (1998) 012,
  \href{http://xxx.lanl.gov/abs/hep-th/9803260}{{\tt hep-th/9803260}}.

\bibitem{Gibbons:1998hm}
G.~W. Gibbons and G.~Papadopoulos, ``Calibrations and intersecting branes,''
  {\em Commun. Math. Phys.} {\bf 202} (1999) 593,
  \href{http://xxx.lanl.gov/abs/hep-th/9803163}{{\tt hep-th/9803163}}.

\bibitem{Gutowski:1999iu}
J.~Gutowski and G.~Papadopoulos, ``Ad{S} calibrations,'' {\em Phys. Lett.} {\bf
  B462} (1999) 81, \href{http://xxx.lanl.gov/abs/hep-th/9902034}{{\tt
  hep-th/9902034}}.

\bibitem{Gutowski:1999tu}
J.~Gutowski, G.~Papadopoulos, and P.~K. Townsend, ``Supersymmetry and
  generalized calibrations,'' {\em Phys. Rev.} {\bf D60} (1999) 106006,
  \href{http://xxx.lanl.gov/abs/hep-th/9905156}{{\tt hep-th/9905156}}.

\bibitem{Gauntlett:2002fz}
J.~P. Gauntlett and S.~Pakis, ``The geometry of {D} = 11 {K}illing spinors.,''
  {\em JHEP} {\bf 04} (2003) 039,
  \href{http://xxx.lanl.gov/abs/hep-th/0212008}{{\tt hep-th/0212008}}.

\bibitem{Figueroa-O'Farrill:2002ft}
J.~Figueroa-O'Farrill and G.~Papadopoulos, ``Maximally supersymmetric solutions
  of ten-dimensional and eleven-dimensional supergravities,'' {\em JHEP} {\bf
  03} (2003) 048, \href{http://xxx.lanl.gov/abs/hep-th/0211089}{{\tt
  hep-th/0211089}}.

\bibitem{Gauntlett:2003fk}
J.~P. Gauntlett and J.~B. Gutowski, ``All supersymmetric solutions of minimal
  gauged supergravity in five dimensions,'' {\em Phys. Rev.} {\bf D68} (2003)
  105009, \href{http://xxx.lanl.gov/abs/hep-th/0304064}{{\tt hep-th/0304064}}.

\bibitem{Fradkin:1985qd}
E.~S. Fradkin and A.~A. Tseytlin, ``Nonlinear {E}lectrodynamics {F}rom
  {Q}uantized {S}trings,'' {\em Phys. Lett.} {\bf B163} (1985) 123.

\bibitem{Leigh:1989jq}
R.~G. Leigh, ``Dirac-{B}orn-{I}nfeld action from {D}irichlet sigma model,''
  {\em Mod. Phys. Lett.} {\bf A4} (1989) 2767.

\bibitem{Witten:1996im}
E.~Witten, ``Bound states of strings and p-branes,'' {\em Nucl. Phys.} {\bf
  B460} (1996) 335--350, \href{http://xxx.lanl.gov/abs/hep-th/9510135}{{\tt
  hep-th/9510135}}.

\bibitem{PrOfAlGe}
P.~Griffiths and J.~Harris, {\em Principles of Algebraic Geometry}.
\newblock Wiley Classics Library, 1994.

\bibitem{Figueroa-O'Farrill:1998su}
J.~M. Figueroa-O'Farrill, ``Intersecting brane geometries,'' {\em J. Geom.
  Phys.} {\bf 35} (2000) 99--125,
  \href{http://xxx.lanl.gov/abs/hep-th/9806040}{{\tt hep-th/9806040}}.

\bibitem{Townsend:1999hi}
P.~K. Townsend, ``Brane theory solitons,'' {\em Cargese 1999, Progress in
  string theory and M-theory} (1999) 265--296,
  \href{http://xxx.lanl.gov/abs/hep-th/0004039}{{\tt hep-th/0004039}}.

\bibitem{Souriau}
J.~M. Souriau, ``Construction explicite de l'indice de {M}aslov,''.

\bibitem{GeometricAsymptotics}
V.~Guillemin and S.~Sternberg, ``Geometric {A}symptotics,'' {\em Mathemaical
  {S}urveys and {M}onographs} {\bf Nr. 14.} (1977).

\bibitem{Woodhouse}
N.~Woodhouse, {\em Geometric {Q}uantization}, vol.~{O}xford {M}athematical
  {M}onographs.
\newblock {O}xford: {C}larendon {P}ress, 1997.

\bibitem{Berkooz:1996km}
M.~Berkooz, M.~Douglas, and R.~Leigh, ``Branes intersecting at angles,'' {\em
  Nucl. Phys.} {\bf B480} (1996) 265--278,
  \href{http://xxx.lanl.gov/abs/hep-th/9606139}{{\tt hep-th/9606139}}.

\bibitem{Ohta:1997fr}
N.~Ohta and P.~K. Townsend, ``Supersymmetry of {M}-branes at angles,'' {\em
  Phys. Lett.} {\bf B418} (1998) 77--84,
  \href{http://xxx.lanl.gov/abs/hep-th/9710129}{{\tt hep-th/9710129}}.

\bibitem{Townsend:1997mk}
P.~K. Townsend, ``M-branes at angles,'' {\em Nucl. Phys. Proc. Suppl.} {\bf 67}
  (1998) 88--92, \href{http://xxx.lanl.gov/abs/hep-th/9708074}{{\tt
  hep-th/9708074}}.

\bibitem{Witten:2000mf}
E.~Witten, ``B{P}{S} bound states of {D}0-{D}6 and {D}0-{D}8 systems in a
  {B}-field,'' {\em JHEP} {\bf 04} (2002) 012,
  \href{http://xxx.lanl.gov/abs/hep-th/0012054}{{\tt hep-th/0012054}}.

\bibitem{Mitra:2000wr}
I.~Mitra and S.~Roy, ``({N}{S}5, {D}p) and ({N}{S}5, {D}(p+2), {D}p) bound
  states of type {I}{I}{B} and type {I}{I}{A} string theories,'' {\em JHEP}
  {\bf 02} (2001) 026, \href{http://xxx.lanl.gov/abs/hep-th/0011236}{{\tt
  hep-th/0011236}}.

\bibitem{Ohta:2001dh}
K.~Ohta, ``Supersymmetric {D}-brane bound states with {B}-field and higher
  dimensional instantons on noncommutative geometry,'' {\em Phys. Rev.} {\bf
  D64} (2001) 046003, \href{http://xxx.lanl.gov/abs/hep-th/0101082}{{\tt
  hep-th/0101082}}.

\bibitem{KLM}
B.~K{\"o}rs, D.~L{\"u}st, and A.~Miemiec, ``Noncommutative {D}-{B}rane and
  {M}-{B}rane {B}ound {S}tates,'' {\em Fortsch. Phys.} {\bf 49} (2001)
  869--884, \href{http://xxx.lanl.gov/abs/hep-th/0103203}{{\tt
  hep-th/0103203}}.

\bibitem{Karch:1998sj}
A.~Karch, D.~L{\"u}st, and A.~Miemiec, ``N = 1 supersymmetric gauge theories
  and supersymmetric 3- cycles,'' {\em Nucl. Phys.} {\bf B553} (1999) 483--510,
  \href{http://xxx.lanl.gov/abs/hep-th/9810254}{{\tt hep-th/9810254}}.

\bibitem{PhD}
A.~Miemiec, ``Branes between geometry and gauge theory,'' {\em Fortsch.Phys.}
  {\bf 48} (2000) 1143--1227.

\bibitem{Miemiec:2005ry}
A.~Miemiec and I.~Schnakenburg, ``Basics of {M}-{T}heory,''
  \href{http://xxx.lanl.gov/abs/hep-th/0509137}{{\tt hep-th/0509137}}.


\end{thebibliography}
\end{document}